\documentclass[twocolumn,showpacs]{revtex4-1}
\usepackage[utf8x]{inputenc}
\usepackage{graphics,epsfig}

\begin{document}
\title{Covariant perturbations through a simple nonsingular bounce}
\author{Atanu Kumar \footnote{atanu.kumar@saha.ac.in}}
\affiliation{Theory Division, Saha Institute of Nuclear Physics,\\
1/AF, Bidhannagar, Kolkata 700098, India\\
}

\newcommand{\fa}{\frac{1}{2}}
\newcommand{\fb}{\frac{1}{3}}
\newcommand{\fc}{\frac{2}{3}}
\newcommand{\fd}{\frac{4}{3}}
\newcommand{\fe}{\frac{3}{4}}
\newcommand{\mc}{\mbox{Curl}}
\newcommand{\lp}{\tilde{\nabla}^2}
\newcommand{\be}{\begin{equation}}
\newcommand{\ee}{\end{equation}}
\newcommand{\ba}{\begin{eqnarray}}
\newcommand{\ea}{\end{eqnarray}}
\newcommand{\bn}{\begin{eqnarray*}}
\newcommand{\en}{\end{eqnarray*}}
\newcommand{\cs}{c_s^2}
\newcommand{\mh}{\mathcal{H}}
\newcommand{\me}{\mathcal{E}}
\newcommand{\mb}{\mathcal{B}}
\newcommand{\muu}{\mathcal{U}}
  \newcommand{\ug}{\muu^{\mbox{GI}}}
\newcommand{\mg}{\delta\mu^{\mbox{GI}}}
\newcommand{\pg}{\delta p^{\mbox{GI}}}
\newcommand{\bt}{\beta}
\newcommand{\xr}{-\frac{1}{\sqrt{3}}}
\newcommand{\xa}{x^2+1}
\newcommand{\xb}{3x^2-1}
\newcommand{\sd}{h_a^{~b}\nabla_b}
\newcommand{\si}{\sigma}
\newcommand{\st}{\sigma^T}
\newcommand{\sv}{\sigma^V}
\newcommand{\sr}{\sigma^S}
\newcommand{\om}{\omega}
\newcommand{\Si}{\Sigma}
\newcommand{\Om}{\Omega}
\newcommand{\ta}{\theta}
\newcommand{\dt}{u^b\nabla_b}
\newcommand{\eo}{\eta_0}
\newcommand{\lt}{\left(}
\newcommand{\rt}{\right)}
\newcommand{\lb}{\left|}
\newcommand{\rb}{\right|}
\newcommand{\Lt}{\left[}
\newcommand{\Rt}{\right]}
\newcommand{\de}{\delta}
\newcommand{\lu}{\mathcal{L}_u}
\newcommand{\ql}{\quad\quad\quad\quad\quad\quad\quad\quad}
\newcommand{\qs}{\quad\quad\quad\quad}
\newcommand{\bX}{\bar{X}^B}
\newcommand{\bY}{\bar{Y}^B}
\newcommand{\bZ}{\bar{Z}^B}
\newcommand{\bA}{\bar{A}^B}
\newcommand{\bg}{\bar{g}}
\newcommand{\dg}{\delta g}
\newcommand{\pa}{\partial}
\newcommand{\bG}{\bar{\Gamma}}
\newcommand{\dG}{\delta{\Gamma}}
\newcommand{\bu}{\bar{u}}
\newcommand{\bmu}{\bar{\mu}}
\newcommand{\bp}{\bar{p}}
\newcommand{\bT}{\bar{T}}
\newcommand{\du}{\delta{u}}
\newcommand{\dm}{\delta{\mu}}
\newcommand{\dep}{\delta{p}}
\newcommand{\dT}{\delta{T}}
\newcommand{\vn}{\vec{\nabla}^2}
\newcommand{\zw}{\delta \mathcal{R}}
\newcommand{\vw}{V^{\mbox{new}}}
\newcommand{\mv}{\mathcal{V}}
\newcommand{\mx}{\mathcal{X}}
\newcommand{\mw}{\mathcal{W}}
\newcommand{\mz}{\mathcal{Z}}
\newcommand{\ex}{x_{\mbox{exit}}}

\begin{abstract}
In this paper we study the evolution of cosmological perturbations through a nonsingular bouncing universe using covariant perturbation theory and examine the validity of linear perturbation theory. The bounce is modeled by a two component perfect fluid. Gauge invariant perturbations are defined in terms of the comoving observers. The scalar and vector perturbations become singular at the turning point, which is the boundary of the spacetime region where the null energy condition is violated. Nonadiabatic modes of comoving curvature perturbation diverge at the turning point. The gravitational waves oscillate around the bounce and the turning point. By computing the growth of linearity parameters, it has been shown that the perturbations do not remain linear at the turning point. We also study the matching conditions on scalar perturbations and $\mv$, related to the spatial curvature perturbation, is found to be the appropriate variable to be matched across transition surface. 
\end{abstract}

\maketitle

\section{Introduction}

Although the standard cosmological model, based on Einstein's general theory of relativity and assumptions of homogeneity and isotropy, has provided a very successful description of the Universe, it suffers from several difficulties. Some of these difficulties have been resolved by assuming an inflationary phase in the early period of expansion of the Universe. However, the initial singularity is a major drawback, because at the singularity, curvature and energy density blow up and therefore, the description of the spacetime in terms of classical physics breaks down. Models of nonsingular universes that have an initial contracting phase followed by a phase of re-expansion after attaining a minimum size (bounce) have been studied for a long time as alternatives to the standard big-bang inflationary models \cite{Mukhanov:1991zn}-\cite{Brandenberger:2012zb}.

Inflation has explained the origin and scale invariance of the spectrum of primordial perturbations. However, it is observed recently that the bounces with a matter dominated contracting phase can also generate scale invariant curvature perturbations \cite{Wands:1998yp,Finelli:2001sr} (after perturbations exit the Hubble radius with a bluish spectrum, contractions boost longer wavelengths more than the shorter wavelengths, thus producing a scale invariant spectrum). 

For being observed in the expanding phase of the Universe, the perturbations must evolve through the bounce \cite{Bozza:2005xs}. But the growing modes of perturbations raise doubts on the validity of linear perturbation theory near bounce \cite{Lyth:2001nv,Battefeld:2004cd} and preservation of scale invariant spectrum of the perturbations. In noncovariant perturbation theory, the perturbations are observed to grow in some gauges while they remain small in some other gauges \cite{Gasperini:2003pb}-\cite{Vitenti:2011yc}. Recently, using covariant perturbation theory it has been shown in \cite{Kumar:2012tr} that in a single fluid dominated contracting branch of a bouncing universe the higher order perturbations grow more rapidly in comparison to the linear order perturbations. However, in order to investigate the behavior of perturbations at the bounce, we need to study a specific model of the nonsingular and bouncing universe. In the new ekpyrotic bouncing model \cite{Khoury:2001wf}, the adiabatic modes of perturbations are observed to be amplified exponentially at the turning points i.e. the boundary of contracting phase and bouncing phase, resulting in breakdown of perturbation theory and spoils of scale invariant spectrum \cite{Xue:2010ux}-\cite{Cai:2013kja}.

From the singularity theorems of Hawking and Penrose \cite{Hawking:1973uf} it is known that a cosmological singularity is unavoidable if the dynamics of a universe is described by classical general relativity and the matter sector obeys some mild energy conditions. To get a bouncing solution one has to either abandon classical general relativity or introduce some unusual matter that violates these energy conditions. In this paper we choose a model that makes use the latter option. 

\begin{figure}
 \begin{center}
  \includegraphics[height=9cm,angle=-90]{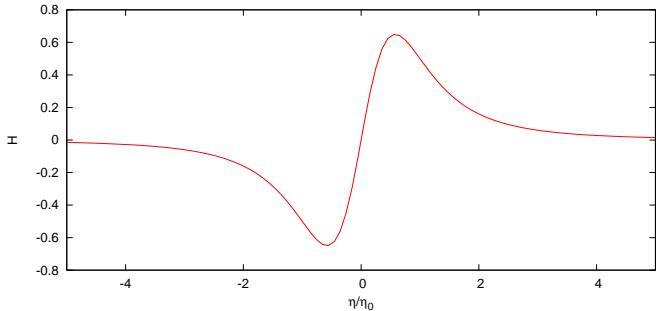}
 \end{center}
\caption{Plot of Hubble parameter as a function of conformal time $\eta$}
\label{hubble}
\end{figure}

We take a toy model for the flat Friedmann-Lemaitre-Robertson-Walker (FLRW) bouncing universe filled with a two-component perfect fluid, one component is a normal fluid with a dustlike equation of state, henceforth referred to as fluid-1, and the other component has a negative energy density and pressure, henceforth referred to as fluid-2 \cite{Finelli:2007tr}. Away from the bounce, the contribution of fluid-2 in the total energy budget is negligible and hence, the contraction of the universe is essentially guided by fluid-1. However, close to the bounce, fluid-2 becomes dominant and as a result the collapse slows down by minimizing the Hubble parameter $H$. At turning point $\dot{H}$ becomes zero. Eventually the bouncing point $H=0$ is reached and the universe starts to re-expand. Again at another turning point $\dot{H}$ vanishes and subsequently fluid-1 starts to dominate. Between the two turning points the null energy condition,
\be
T_{\mu\nu}k^{\mu}k^{\nu}\ge 0 \quad \text{for any null vector $k^{\mu}$},
\ee
is violated by the composite fluid. Variation of Hubbile parameter as function of conformal time $\eta$ is shown in \figurename{(\ref{hubble})}.

In this paper we study the evolution of perturbations through the bounce in the covariant approach. It turns out that the scalar and vector perturbations diverge not at the bouncing point but at the turning point; whereas the tensor perturbations oscillate at the bounce as well as at the turning point. At the turning point, we investigate the validity of linear perturbation theory. The linearity parameters (the ratio of the nonlinear and linear terms in perturbation equations) diverge at the turning point, confirming the appearance of nonlinearity in perturbations. The comoving curvature perturbation is conserved for adiabatic perturbations. However, in our model a nonadiabatic mode of perturbation exists. We have computed the nonadiabatic mode of covariantly defined comoving curvature perturbation and have shown that that mode is singular at the turning point. 

We also consider a specific initial condition for scalars in which the entropic perturbation is absent and the adiabatic perturbations are originated from quantum fluctuations of the Bunch-Davis vacuum state in the matter dominated era. Using a numerical analysis we evolve the perturbations through the bounce. Divergence of the linearity parameters remains unaltered even in the presence of these special initial conditions. The scale invariance of the spectra are preserved well after the bounce. The correct spectra are obtained from the matching of $\mv$ and not the $\mx$ across the transition surface.  

The paper is organized as follows. In Sec.~\ref{background}, we describe the background bouncing model. In Sec.~\ref{perturbations} the gauge invariant perturbations are defined covariantly. The equations are set up in Sec.~\ref{equations}. In Sec.~\ref{solutions} we demonstrate the solutions of linear perturbation equations. In Sec.~\ref{curvature}, the behavior of comoving curvature perturbations are discussed. In Sec.~\ref{compare}, we compute the linearity parameters at the turning point. The matching conditions are discussed in Sec.~\ref{matching} and the numerical analysis is demonstrated in Sec.~\ref{numerical}. In \appendixname{~\ref{harmonics}} we describe the spatial harmonics on 3-hypersurface and in \appendixname{~\ref{ordinary}} the covariant perturbations are expressed in terms of ordinary gauge invariant variables of coordinate based perturbation theory.


\section{Background}
\label{background}

We consider a flat FLRW universe with a two component perfect fluid \cite{Finelli:2007tr}. The two components have the same 4-velocity $u^a$ which is taken to be the velocity of the comoving observers,
\be
u^a=\frac{dx^a}{d\tau}, \quad u^au_a=-1,
\ee
where $\tau$ is the proper time along the world lines of comoving observers. The two components of the fluid must have identical velocity at least in the background spacetime, because otherwise the background ceases to be an isotropic one. We assume here that the fluids' velocity is the same in the physical spacetime also. Although this assumption may lead to some loss of accuracy, our aim in this paper is not to calculate the cosmological parameters accurately but to understand the physical consequences of bounce on the evolution of perturbations. We hope such an assumption does not significantly alter the qualitative results. Note that such assumptions are often taken into consideration for matter-radiation transition in expanding universes \cite{Weinberg:2008zzc}.

The dynamical evolution is determined by the Einstein equation,
\be
G_{ab}=R_{ab}-\fa g_{ab}R=\kappa T_{ab},
\ee
where $\kappa=8\pi G$. $T_{ab}$ is the total energy-momentum tensor,
\be
T_{ab}=\mu u_au_b + p h_{ab}, \quad h_{ab}=g_{ab}+u_au_b,
\ee
\be
\mu=\mu_1-\mu_2, \quad p=p_1-p_2
\ee
Fluid-1 is a normal fluid, whereas fluid-2 violates the strong and weak energy condition. Each component satisfies the energy conservation condition separately:
\be
\label{energy} \dot{\mu}_1+\ta(\mu_1+p_1)=0, \quad \dot{\mu}_2+\ta(\mu_2+p_2)=0.
\ee
the overdot is representing the covariant derivative along world lines of comoving observers and $\ta=\nabla_au^a$ is the expansion of neighboring world lines of comoving observers. We can define a scale $a(\tau)$ along each world line as
\begin{equation}
 \label{theta} \ta = 3\frac{\dot{a}}{a},
\end{equation}
where $a$ can be determined up to a multiplicative constant and is identified as the Robertson-Walker scale factor.

Equation (\ref{energy}) together with the equations of state, $p_1=w_1\mu_1$, $p_2=w_2\mu_2$, give the evolution of energy densities,
\be
\mu_1=\frac{M_1}{a^{n_1}}, \quad \mu_2=\frac{M_2}{a^{n_2}},
\ee
where $n_1=3(1+w_1)$ and $n_2=3(1+w_2)$.

If $w_1$ and $w_2$ bears the following relation,
\be
\label{wrel} w_2=2w_1+\fb \Leftrightarrow n_2=2(n_1-1),
\ee
then the Friedmann's equations yield a simple bouncing solution,
\be
a(\eta)=\epsilon\left( 1+\frac{\eta^2}{\eta_0^2} \right)^{\alpha},
\ee
where $\eta=\int a^{-1}dt$ is conformal time and,
\be
\epsilon=\left(\frac{M_2}{M_1}\right)^\alpha, \quad \alpha=\frac{1}{n_2-n_1}=\frac{1}{n_1-2}
\ee

At any point on the manifold, a perfect fluid is characterized completely by energy density $\mu$, entropy density $S$ and the velocity 4-vector $u_a$. The pressure can be expressed as a function of $\mu$ and $S$ via equation of state
\be
p=p(\mu,S).
\ee 
So the small change in pressure is given by
\be
\de p=\cs\de\mu+\tau\de S,
\ee
where $\cs=\lt\frac{\partial p}{\partial \mu}\rt_S$ is adiabatic speed of sound and $\tau=\lt\frac{\partial p}{\partial S}\rt_{\mu}$. Since in absence of dissipation, entropy is conserved along fluid flow lines, i.e. $\dot{S}=0$,
\be
\cs=\frac{\dot{p}}{\dot{\mu}}=-\frac{\dot{p}}{\ta(\mu+p)}
\ee

This shows that if $\mu+p$ vanishes, but $\dot{p}$ remains nonzero, then the speed of sound blows up.

Let us consider the normal fluid is dustlike, i. e. $w_1=0$. Then the relation (\ref{wrel}) constrains the fluid-2 to be radiationlike ($w_2=\fb$):
\be
\mu_1=\frac{M_1}{a^{3}}, \quad \mu_2=\frac{M_2}{a^{4}}
\ee

In terms of the dimensionless quantity $x=\eta/\eo$, we have
\ba
a(x)&=&\epsilon(1+x^2), \quad \epsilon=\frac{M_2}{M_1}, \quad \kappa M_1\eta_0^2=12\epsilon, \\
\mh&=&\frac{a'}{a}=\frac{2x}{1+x^2}, \quad \mh'=2\frac{1-x^2}{1+x^2}, \\ 
\frac{a''}{a}&=&\mh'+\mh^2=\frac{2}{1+x^2}.
\ea
Primes are representing derivatives with respect to $x$.

The scalar curvature $R=\frac{6}{\eta_0^2a^2}\frac{a''}{a}$ remains finite for the entire range of $x$:

\be
\label{muplusp} \mu+p =\frac{M_1}{a^{4}}(a-\bt)=\frac{M_1}{\epsilon^3}\frac{3x^2-1}{3(x^2+1)^4}, \quad \bt=\fd\epsilon.
\ee
So, at $x=\pm\frac{1}{\sqrt{3}}$, i.e. $a=\bt$, $\mu+p$ vanishes. The null energy condition, which in the case of perfect fluid means $\mu+p\ge0$, is satisfied for $|x|\ge\frac{1}{\sqrt{3}}$, but it is violated for $|x|<\xr$. The spacelike hypersurfaces at $x=\pm\frac{1}{\sqrt{3}}$, which form the boundary between the two regions, are called turning points. 

Speed of sound in this model diverges at the turning points,
\be
\cs=-\fb\frac{\bt}{a-\bt}=-\fd\frac{1}{3x^2-1}
\ee


\section{Perturbations}
\label{perturbations}

In covariant perturbation theory, as gauge invariant perturbations, we consider the variables, which vanish in the background FLRW manifold. Some of those variables which form a closed set of equations are listed below \cite{Dunsby:1991xk}:

(1) Shear, vorticity and acceleration,
 \ba
\label{shear}  \si_{ab} &=& (h_{(a}^{~c}h_{b)}^{~d}-\frac{1}{3}h_{ab}h^{cd})\nabla_d u_c, \\
  \om_{ab} &=& h_{[a}^{~c}h_{b]}^{~d}\nabla_d u_c, \\
  \nu_a &=& u^b\nabla_b u_a.
 \ea
Rotation vector is defined as $\omega_a=\frac{1}{2}\epsilon_{abc}\omega^{bc}$. $\epsilon_{abc}$ is Levi-Civita tensor in the 3-hypersurface
defined by $\epsilon_{abc}=\eta_{abcd}u^d$. \\
(2) ``Electric'' and ``magnetic'' parts of the Weyl tensor,
   \ba
    E_{ab}=C_{acbd}u^cu^d, \quad  H_{ab}=\frac{1}{2}C_{acpq}\eta^{pq}_{~~bd}u^cu^d.
   \ea
(3) Spatial gradients of the energy densities, pressure densities and expansion,
\ba
X_{a}&=&\kappa\sd\mu, \quad Y_{a}=\kappa\sd p, \quad Z_a=\sd\ta,\nonumber \\
X_{1a}&=&\kappa\sd\mu_1, \quad X_{2a}=\kappa\sd\mu_2, \nonumber\\
Y_{1a}&=&\kappa\sd p_1, \quad Y_{2a}=\kappa\sd p_2,\nonumber\\
X_a &=& X_{1a}-X_{2a}, \quad Y_a=Y_{1a}-Y_{2a}.
\ea
Using equations of state, $Y_{1a}=0, Y_{2a}=\fb X_{2a}$.

The nonadiabatic mode of perturbation is defined as
\ba
\label{entropy} \Gamma_a=\kappa\tau\sd S=Y_a-\cs X_a=\frac{\bt X_{1a}-aX_{2a}}{3(a-\bt)}.
\ea

All variables defined in (\ref{shear})-(\ref{entropy}) and their derivatives are considered to be linear or first order variables. These first order variables can be translated to the ordinary gauge invariant perturbations used in coordinate based perturbation theory. Some of those relations are shown in \appendixname{~\ref{ordinary}}. Any quantity which is quadratic in first order variables is said to be nonlinear.


\section{Dynamic equations and constraints}
\label{equations}

We assume the two components do not exchange energy but exchange momentum among themselves. So the momentum conservation equation must be satisfied for the two fluids together:
\be
\label{momentum}\kappa(\mu+p)\nu_a+Y_a=0.
\ee    

Taking the spatial derivative of the equations in (\ref{energy}), we obtain
\ba
\label{x1dot}  & &a^{-4}h_a^{~b}(a^4X_{1a}\dot{)} = \ta\left(\frac{\mu_2+p_2}{\mu+p}Y_{1a}-\frac{\mu_1+p_1}{\mu+p}Y_{2a}\right) \nonumber \\ 
&& \qs\quad -\kappa(\mu_1+p_1)Z_a -(\si^b_{~a}+\om^b_{~a})X_{1b},   \\
\label{x2dot} && a^{-4}h_a^{~b}(a^4X_{2a}\dot{)} = \ta\left(\frac{\mu_2+p_2}{\mu+p}Y_{1a}-\frac{\mu_1+p_1}{\mu+p}Y_{2a} \right) \nonumber \\
&& \qs\quad -\kappa(\mu_2+p_2)Z_a -(\si^b_{~a}+\om^b_{~a})X_{2b}. 
\ea

Subtracting (\ref{x2dot}) from (\ref{x1dot}),
\ba
\label{xdot} a^{-4}h_a^{~b}(a^4X_{a}\dot{)} = -\kappa(\mu+p)Z_a-(\si^b_{~a}+\om^b_{~a})X_{b}.
\ea

Other equations are \cite{Hawking,Ellis:1989jt}
\ba
a^{-3}h_{a}^{~b}(a^3Z_b\dot{)}&=& \mathcal{R}\nu_a - \fa X_a + A_a + 2h_{a}^{~b}\nabla_b(\omega^2-\sigma^2) \nonumber \\
\label{zdot} & & \qs - (\omega^b_{~a}+\sigma^b_{~a})Z_b,  
\ea
where
\be
\label{r} \mathcal{R} = \kappa\mu -\fb\theta^2+\nabla^a\nu_a+2(\omega^2-\sigma^2).
\ee
and 
\be
A_a=\sd \nabla^c\nu_c
\ee

\ba
\label{odot} & a^{-2}h_{a}^{~c}h_{b}^{~d}(a^2\omega_{cd}\dot{)} = h_{a}^{~c}h_{b}^{~d}\nabla_{[d}\nu_{c]} + 2\sigma_{c[a}\omega_{b]}^{~c}, \\
\label{sdot} & a^{-2}h_{a}^{~c}h_{b}^{~d}(a^2\sigma_{cd}\dot{)} = -E_{ab} + \nabla_{\langle b}\nu_{a\rangle} - \omega_{ac}\omega^c_{~b} -\sigma_{ac}\sigma^c_{~b} \nonumber \\
&  +\frac{2}{3}h_{ab}(\sigma^2-\omega^2)+\nu_a\nu_b, \\
\label{edot} & a^{-3}h_{a}^{~c}h_{b}^{~d}(a^3E_{cd}\dot{)} = - \mc H_{ab} -\fa\kappa(\mu+p)\sigma_{ab} + E^c_{~(a}\omega_{b)c} \nonumber \\
& + E^c_{~(a}\sigma_{b)c} + \epsilon_{acd}\epsilon_{bpq}\sigma^{cp}E^{dq} - 2H^c_{~(a}\epsilon_{b)cd}\nu^d, \\
\label{hdot} & a^{-3}h_{a}^{~c}h_{b}^{~d}(a^3H_{cd}\dot{)} = \mc E_{ab} + H^c_{~(a}\omega_{b)c} + H^c_{~(a}\sigma_{b)c} \nonumber \\
 &  +\epsilon_{acd}\epsilon_{bpq}\sigma^{cp}H^{dq}  - 2H^c_{~(a}\epsilon_{b)cd}\nu^d .
\ea

We have used following notations : 
\bn
 & &\lambda_{(ab)} = \fa(\lambda_{ab}+\lambda_{ba}), \quad \lambda_{[ab]}=\fa(\lambda_{ab}-\lambda_{ba}), \\
 & & \lambda_{\langle ab\rangle} = h_{a}^{~c}h_{b}^{~d}(\lambda_{(cd)}-\fb h_{cd}\lambda^e_{~e}), \quad \mc \lambda_{ab} = h^e_{~(a}\epsilon_{b)cd}\nabla^d\lambda_e^{~c}. \\
 & & 
\en

There are also constraint relations which must be satisfied at some initial time on each world line ,
\ba
\label{delom} & & h_{a}^{~c}\nabla_b(\omega^b_{~c}+\sigma^b_{~c})-\nu^b(\omega_{ab}+\sigma_{ab}) = \frac{2}{3}Z_a, \\
\label{delomega} & & \nabla_a\omega^a = 2\nu_a\omega^a, \\
& & \mc\omega_{ab}+\mc\sigma_{ab} = -H_{ab}, \\
& & h_{a}^{~c}\nabla^bE_{bc}+3H_{ab}\omega^b-\epsilon_{abc}\sigma^b_{~d}H^{cd} = \fb X_a, \\
& & h_{a}^{~c}\nabla^bH_{bc}-3E_{ab}\omega^b-\epsilon_{abc}\sigma^b_{~d}H^{cd} = \kappa(\mu+p) \omega_a.
\ea


\section{Solutions of linearized equations}
\label{solutions}
To study the linear evolution of perturbations we will use the usual classification of perturbations in terms of scalar, vector and tensor modes. $X_a$s, $Y_a$s, $Z_a$, $\Gamma_a$, and $\nu_a$, constructed from spatial gradients of scalar functions, are considered as scalar perturbations. The antisymmetric tensor $\om_{ab}$ is dual to a vector $\om_a$. By Eq. (\ref{delomega}), the $\om_a$ is a divergence-less vector up to first order. So we consider $\om_{ab}$ as a pure vector perturbation. The $\si_{ab}$, $E_{ab}$ and $H_{ab}$ are traceless symmetric tensors. These variables can be decomposed into scalar, vector and pure tensor perturbation.      

\subsection{Scalar perturbations}

Linearized forms of (\ref{x1dot})-(\ref{zdot}) in our background model are
\ba
\label{x1dotl} a^{-4}(a^4X_{1a}\dot{)} &=& -\frac{\kappa M_1}{a^3}Z_a-\fb\ta\frac{a}{a-\bt}X_{2a}, \\
\label{x2dotl} a^{-4}(a^4X_{2a}\dot{)} &=& -\frac{\bt\kappa M_1}{a^4}Z_a-\fb\ta\frac{a}{a-\bt}X_{2a}, \\
\label{zdotl} a^{-3}(a^3Z_a\dot{)} &=& -\fa(X_{1a}-X_{2a})+A_a.
\ea

To solve the equations, let us expand the variables in Fourier modes on the 3-hypersurface,
\be
\label{sfourier} S_{a} = \sum_k S(k,t)Q^{(0)}_{a},
\ee
where $S$ stands for any scalar perturbations. $Q^{(0)}_{a}$ are the eigenfunctions of spatial Laplacian, explained in the \appendixname{~\ref{harmonics}}.

Using (\ref{momentum}),
\ba
A_a &=& \sd\nabla^c\nu_c \nonumber \\
&=& \frac{1}{\kappa M_1}\frac{a^4}{a-\bt}\sum_k \frac{k^2}{a^2} Y(k,t)Q^{(0)}_{a} \nonumber \\
&=& -\frac{1}{3\kappa M_1}\frac{a^4}{a-\bt}\sum_k \frac{k^2}{a^2} X_2(k,t)Q^{(0)}_{a}
\ea

Using dimensionless quantities,
\ba
\mathcal{X}_1=\eo^3a^4X_1, \quad \mathcal{X}_2=\eo^3a^4X_2, \quad \mathcal{Z}=\eo^2a^3Z,
\ea
Eqs. (\ref{x1dotl})-(\ref{zdotl}) in Fourier modes become
\ba
\label{x1pri} \mathcal{X}_1'&=&-\frac{9\bt}{a}\mathcal{Z}-\frac{a'}{a-\bt}\mathcal{X}_2, \\
\label{x2pri} \mathcal{X}_2'&=&-\frac{9\bt^2}{a^2}\mathcal{Z}-\frac{a'}{a-\bt}\mathcal{X}_2,\\
\label{zpri} \mathcal{Z}'&=&-\fa(\mathcal{X}_1-\mathcal{X}_2)-\frac{q^2a^2}{27\bt(a-\bt)}\mathcal{X}_2,
\ea
where prime denotes the derivative with respect to dimensionless conformal time $x=\eta/\eo$ and $q=k\eo$ is the dimensionless wave number.

Eliminating $\mathcal{Z}$ from (\ref{x1pri}) and (\ref{x2pri}),
\be
\bt \mathcal{X}_1'=(a\mathcal{X}_2)'\quad \Longrightarrow \mathcal{X}_2(q,x)=\frac{\bt}{a}(\mathcal{X}_1(q,x)-C_1(q))
\ee

Using new variable $\mathcal{W}=\mathcal{X}_1-C_1$, Eqs. (\ref{x1pri})-(\ref{zpri}) reduce to
\ba
\label{wprime} \mathcal{W}'&=&-\frac{a'}{a}\frac{\bt}{a-\bt}\mathcal{W}-\frac{9\bt}{a}\mathcal{Z}, \\
\label{zprime} \mathcal{Z}'&=&-\left( \fa\frac{a-\bt}{a}+\frac{q^2}{27}\frac{a}{a-\bt} \right)\mathcal{W}-\fa C_1.
\ea

The arbitrary constant $C_1(q)$ is related to the initial spectrum of nonadiabatic mode of perturbation, defined in (\ref{entropy}),
\ba
\label{Gamma_qx}\Gamma(q,x)=\frac{\bt C_1(q)}{3\eo^3a^4(a-\bt)}.
\ea
This shows that entropy perturbation decays far away from bounce ($a\gg\bt$) as $a^{-5}$ and diverges at $a\sim \bt$.

From (\ref{wprime}) and (\ref{zprime}) we extract a second order inhomogeneous differential equation for $\mathcal{W}$: 

\begin{widetext}

\ba
\mathcal{W}'' + \frac{a'}{a-\bt}\mathcal{W}' + \frac{\bt}{a}\left( \frac{a''}{a}-\frac{a'^2}{(a-\bt)^2}-\frac{9}{2}\frac{a-\bt}{a}-\frac{q^2}{3}\frac{a}{a-\bt} \right)\mathcal{W}=\frac{9\bt C_1(q)}{2a}
\ea

Or
\be
\label{udprime} \mathcal{W}''+\frac{6x}{\xb}\mathcal{W}'-\left( \frac{2(9x^2+1)(3x^4-2x^2+3)}{(\xa)^2(\xb)^2}+\fd\frac{q^2}{\xb} \right)\mathcal{W}=\frac{6C_1(q)}{\xa}
\ee

\end{widetext}

$\mathcal{X}_1$, $\mathcal{X}_2$ and $\mathcal{Z}$ can be expressed in terms of $\mathcal{W}$,
\ba
& & \mathcal{X}_1=\mathcal{W}+C_1, \quad \mathcal{X}_2=\frac{\bt}{a}\mathcal{W}=\frac{4}{3(\xa)}\mathcal{W}, \nonumber \\
& & \mathcal{Z}=-\frac{a}{9\bt}\lt \mathcal{W}''+\frac{a'}{a}\frac{a}{a-\bt}\mathcal{W} \rt=-\frac{1}{9}\frac{a^2}{\bt(a-\bt)}\lt \frac{a-\bt}{a}\mathcal{W} \rt'. \nonumber  
\ea

All scalar perturbations are given by
\ba
& X_1(q,x) = \eo^{-3}a^{-4}\mathcal{X}_1(q,x), \quad
X_2(q,x) = \eo^{-3}a^{-4}\mathcal{X}_2(q,x), \nonumber \\
& Z(q,x) = \eo^{-2}a^{-3}\mathcal{Z}(q,x), \quad
Y(q,x) = -\fb X_2(q,x), \nonumber \\
\label{allscalar} & \nu(q,x) = -\frac{Y(q,x)}{\kappa(\mu+p)}, \quad
A(q,x) = -\lt\frac{k}{a}\rt^2\nu(q,x). 
\ea

The shear also has a scalar part, which is obtained from the constraint (\ref{delom}). Linearizing (\ref{delom}), 
\ba
\label{lindelom}\nabla_b(\omega^b_{~a}+\sigma^b_{~a}) = \frac{2}{3}Z_a.
\ea

Now $\si_{ab}$ is a traceless symmetric tensor. Decomposing it into a divergenceless tensor $\st_{ab}$, gradient of a divergenceless vector $\sv_{ab}$ and double gradient of a scalar $\sr_{ab}$ ,
\ba
\si_{ab}=\st_{ab}+\sv_{ab}+\sr_{ab} = \sum_k\st(k,t)Q^{(2)}_{ab} \nonumber \\
+\sum_k\sv(k,t)Q^{(1)}_{ab}+\sum_k\sr(k,t)Q^{(0)}_{ab},~~
\ea

\ba
\nabla_b\si^b_{~a} &=& \nabla^b\sv_{ab} + \nabla^b\sr_{ab}, \nonumber \\
\label{sisv} &=& \sum_k\fa\frac{k}{a}\sv Q^{(1)}_a+\sum_k\fc\frac{k}{a}\sr Q^{(0)}_a.
\ea

Let us define
\ba
\label{omr}  r_a = \nabla^b\om_{ab} = -\nabla_b\om^b_{~a}.
\ea

$r_a$ is a divergence-less vector since
\bn
\nabla^ar_a=\nabla^a\nabla^b\om_{ab}=\fa\lt R^{ab~c}_{~~a}\om_{cb}+R^{ab~c}_{~~b}\om_{ac} \rt=0.
\en

So $r_a$ can written as
\be
\label{rfourier} r_a = \sum_k r(k,t) Q^{(1)}_a.
\ee

So, putting (\ref{sfourier}) and (\ref{sisv})-(\ref{rfourier}) in the constraint (\ref{lindelom}) and separating the scalar and vector parts we obtain
\ba
\label{siscalar} \sr(k,t)=Z(k,t)\frac{a(t)}{k}, \\ 
\label{sivector} \sv(k,t)=2r(k,t)\frac{a(t)}{k}.
\ea

Using (\ref{allscalar}),
\ba
\label{srsol} \sr(q,x)=Z(q,x)\frac{a}{k}=\eo^{-1}a^{-2}\frac{\mathcal{Z}(q,x)}{q}.
\ea

Let us use the opportunity to clarify the mistakes in calculations in Eqs. (93), (94), (101) and (102) of  \cite{Kumar:2012tr}. According to (\ref{siscalar}) and (\ref{sivector}), $\sv$ and $\sr$ in a radiation dominated background behave as
\ba
\sv = \frac{2R(k)}{k}a^{-1}, \\
\sr = \frac{Z^{(1)}(k)}{k}a^{-3} + \frac{Z^{(1)}(k)}{k}.
\ea

In the dust dominated case,
\ba
\sv = \frac{2R(k)}{k}a^{-2}, \\
\sr = \frac{Z^{(1)}(k)}{k}a^{-3} + \frac{Z^{(1)}(k)}{k}a^{-1/2}.
\ea

This is reflected in errors in the calculation of $\varepsilon_2$, $\varepsilon_5$ and $\varepsilon_7$ of the Sec. VI B of \cite{Kumar:2012tr} . The correct form of these parameters follows:
\ba
     \varepsilon_2 = \frac{\left|\Sigma^b_{~a}\bar{X}_b\right|}{\left|\kappa M\bar{Z}_a\right|}a^{-\frac{3}{2}}, \quad
     \varepsilon_5 = \frac{\left|2\Sigma_a\right|}{\left|\frac{1}{2} \bar{X}_a\right|}a^{-\frac{3}{2}}, \quad \nonumber \\
     \varepsilon_7 = \frac{\left|\Sigma^b_{~a}\bar{Z}_b\right|}{\left|\frac{1}{2} \bar{X}_a\right|}a^{-\frac{3}{2}}.
\ea

For $q=0$, Eq. (\ref{udprime}) has a general solution,
\ba
 \mathcal{W}(0,x) &=& -C_1(0)\frac{3(\xa)}{\xb}+C_2(0)\frac{3x}{(\xa)(\xb)}  \nonumber \\
& & \quad +C_3(0)\frac{9x^6+25x^4+15x^2+15}{3(\xa)(\xb)}
\ea

To solve for modes with nonzero momentum, we will concentrate on different regions of interest. We are working in a collapsing FLRW universe undergoing a nonsingular bounce. Long before the bounce ($a\gg\epsilon$), the energy density of fluid-1 dominates over the energy density of fluid-2 and we have a dust dominated collapsing FLRW background. Let us call this region as region A. The neighborhood of the turning point $x\sim \xr$ is region B. Another region of interest is the point of bounce, characterized by vanishing of the Hubble parameter and corresponds to the time $x=0$. This is the region C. \\

\vspace{10mm}
\textbf{Region A}: 

In this region, $|x|\gg1$ and $a(x)\simeq\fe\bt x^2$. Changing the variable $x$ to $z=\frac{1}{x}$, Eq. (\ref{udprime}) takes form
\be
\label{uzz} \frac{d^2\mathcal{W}_A}{dz^2}+P_{A1}(q,z)\frac{d\mathcal{W}_A}{dz}+P_{A0}(q,z)\mathcal{W}_A=C_1(q)P_A(q,z),
\ee
where the coefficients $P_{A1}$, $P_{A0}$, $P_{A}$ are expanded in Taylor series around $z=0$:
\ba
P_{A1}(q,z) &=& -\fc\lt z+\frac{z^3}{3}+\frac{z^5}{9}+\cdots \rt, \nonumber\\
P_{A0}(q,z) &=& -\lt 6+\frac{4q^2}{9} \rt\frac{1}{z^2}+\lt \frac{34}{3}-\frac{4q^2}{27} \rt \nonumber \\
& & \qs - \lt 22+\frac{4q^2}{81} \rt z^2 +\cdots, \nonumber\\
P_A(q,z) &=& 6\lt \frac{1}{z^2}-1+z^2-\cdots \rt.
\ea

The power series solution of (\ref{uzz}) is
\ba
 \mathcal{W}_A(q,\frac{1}{z}) &=& -\frac{C_1(q)}{1+\frac{2q^2}{27}}\lt 1+\fd\frac{9-q^2}{9+q^2}z^2+\cdots \rt \nonumber \\
& &  +   C_2^A(q)z^{3+\de}\lt 1-\frac{28-7\de-\de^2}{6(7+2\de)}z^2+\cdots \rt  \nonumber \\
& &  +   \frac{C_3^A(q)}{z^{2+\de}}\lt 1+\frac{38-3\de-\de^2}{6(3+2\de)}z^2+\cdots \rt, \;\;  
\ea
where
\be
\de=\frac{5}{2}\lt \sqrt{1+\lt\frac{4q}{15}\rt^2}-1 \rt.
\ee

In the limit $z\rightarrow 0$, evaluating the variables,
\ba
\label{u1Aqx} \mathcal{X}_{1A}(q,x) &=& \frac{2q^2}{27+2q^2}C_1(q) + C_2^A(q)x^{-3-\de} \nonumber \\
 & & \ql + C_3^A(q)x^{2+\de,} \nonumber\\
\label{u2Aqx} \mathcal{X}_{2A}(q,x) &=& -\frac{36}{27+2q^2}C_1(q)x^{-2} + \fd C_2^A(q)x^{-5-\de} \nonumber \\
& & \ql + \fd C_3^A(q)x^{\de}, \nonumber\\
\mathcal{Z}_A(q,x) &=& \frac{12q^2}{(27+2q^2)(9+q^2)}\frac{C_1(q)}{x} + \frac{3+\de}{12} \frac{C_2^A(q)}{x^{2+\de}} \nonumber \\
\label{vAqx} & & \qs -\frac{2+\de}{12} C_3^A(q)x^{3+\de}. 
\ea

\vspace{10mm}
\textbf{Region B}: 

In this region, $x\sim\xr$. In terms of a new variable, $y=\sqrt{3}x+1$, (\ref{udprime}) takes the following form:
\be
\label{uyy} \frac{d^2\mathcal{W}_B}{dy^2}+P_{B1}(q,y)\frac{d\mathcal{W}_B}{dy}+P_{B0}(q,y)\mathcal{W}_B=C_1(q)P_B(q,y)
\ee
Again the coefficients obtained as a Taylor series around $y=0$,
\ba
P_{B1}(q,y) &=& \frac{1}{y}\Lt 1-\fa y-\frac{1}{4}y^2+\cdots \Rt, \nonumber\\
P_{B0}(q,y) &=& -\frac{1}{y^2}\Lt 1+\lt\fa-\frac{2q^2}{9}\rt y+\lt\frac{1}{4}-\frac{q^2}{9}\rt y^2 +\cdots \Rt, \nonumber\\
P_B(q,y) &=& \frac{3}{2}\Lt 1+\fa y +\cdots \Rt.
\ea

The general solution of (\ref{uyy}) in the limit $y\rightarrow 0$ is
\ba
\label{wBqy} \mathcal{W}_B(q,y) &=& \fa C_1(q)y^2\Lt 1+\frac{1}{8}\lt 3-\frac{2q^2}{9}\rt y +\cdots \Rt \qs\nonumber \\
&& \; +C_2^B(q)y\Lt 1+\fb\lt 1-\frac{2q^2}{9} \rt y+\cdots \Rt \nonumber \\
&& \; + \frac{C_3^B(q)}{y}\Lt 1+\frac{2q^2}{9}y+\cdots \Rt.,
\ea

\ba
\label{x1Bqy} \mathcal{X}_{1B}(q,y) &=& C_1(q)+C_2^B(q)y+\frac{C_3^B(q)}{y}, \\
\label{x2Bqy} \mathcal{X}_{2B}(q,y) &=& \fa C_1(q)y^2+C_2^B(q)y+\frac{C_3^B(q)}{y},  \\
 \mathcal{Z}_B(q,y) &=& \xr \lt \fa C_1(q)y+ \fc C_2^B(q) + \frac{2q^2}{27}\frac{C_3^B(q)}{y} \rt. \nonumber \\
\label{zBqy} && \ql
\ea

So the scalar perturbations diverge as $y^{-1}$ at the turning point. Though both $\mathcal{X}_1$ and $\mathcal{X}_2$ diverge as $y^{-1}$ near the turning point, the combination $\mx=\mx_1-\mx_2$ remains finite. So $X(q,x) = \eo^{-3}a^{-4}\mx$ is also finite and well behaved at the turning point.

\vspace{10mm}

\textbf{Region C}: $x\sim 0$

\be
\label{uxx} \frac{d^2\mathcal{W}_C}{dx^2}+P_{C1}(q,x)\frac{d\mathcal{W}_C}{dx}+P_{C0}(q,x)\mathcal{W}_C=C_1(q)P_C(q,x),
\ee

\ba
P_{C1}(q,x) &=& -6x(1+3x^2+9x^4+\cdots), \nonumber\\
P_{C0}(q,x) &=& -\lt 6-\frac{4q^2}{3}\rt - \lt 74-4q^2 \rt x^2 \nonumber \\
& & \qs - \lt 278-12q^2 \rt x^4 - \cdots, \nonumber\\
P_C(q,x) &=& 6(1-x^2+x^4-\cdots).
\ea

Solutions:

\begin{widetext}
\ba
\mathcal{W}_C(q,x) &=& C_1(q)\Lt 3x^2+\lt 4-\frac{q^2}{3}\rt x^4 +\cdots \Rt + C_2^C(q)\Lt x+2\lt 1-\frac{q^2}{9} \rt x^3 +\cdots \Rt \nonumber \\
& & \ql\qs + C_3^C(q)\Lt 1+\lt 3-\frac{2q^2}{3}\rt x^2 + \fb\lt 32-5q^2+\frac{2q^4}{9}\rt x^4+\cdots \Rt, \\
\mathcal{X}_{1C}(q,x) &=& C_1(q)\Lt 1+3x^2+\cdots \Rt + C_2^C(q)\Lt x+2\lt 1-\frac{q^2}{9} \rt x^3 +\cdots \Rt + C_3^C(q)\Lt 1+\lt 3-\frac{2q^2}{3}\rt x^2 +\cdots \Rt, \nonumber \\
\mathcal{X}_{2C}(q,x) &=& \fd  C_1(q)\Lt 3x^2+\lt 1-\frac{q^2}{3}\rt x^2+\cdots \Rt + \fd C_2^C(q)\Lt x+\lt 1-\frac{2q^2}{9}\rt x^3+\cdots \Rt \nonumber \\
& & \ql\ql\ql + \fd C_3^C(q)\Lt 1+2\lt 1-\frac{q^2}{3}\rt x^2+\cdots \Rt,  \nonumber \\
\mathcal{Z}_C(q,x) &=& -\frac{1}{6}C_1(q)\Lt 3x -\lt 1+\frac{2q^2}{3}\rt x^3 \Rt -\frac{1}{12}C_2^C(q)\Lt 1-\lt 1+\frac{2q^2}{3}\rt x^2+\cdots \Rt \nonumber \\
& &   \ql\qs  +\frac{1}{6}C_3^C(q)\Lt\lt 1+\frac{2q^2}{3}\rt x + \fb\lt 47+4q^2-\frac{4q^4}{9}\rt x^3+\cdots  \Rt.
\ea
\end{widetext}

So at the point of bounce, scalar perturbations remain finite and well behaved.

\vspace{10mm} 


\subsection{Vector perturbations}

Let us consider the right hand side of (\ref{odot}) up to linear order,
\bn
 \nabla_d\nu_c &=& -\frac{1}{\mu+p}\nabla_d\lt h_c^{~e}\nabla_e p\rt  \\
 &=& -\frac{1}{\mu+p} \Lt h_c^{~e}\nabla_d\nabla_e p +(\nabla_e p)(u_c\nabla_du^e+u^e\nabla_du_c) \Rt 
\en
\ba
 \Rightarrow h_a^{~c}h_b^{~d}\nabla_{[d}\nu_{c]} = -\frac{\dot{p}}{\mu+p} \om_{ab} = \cs\ta\om_{ab}. 
\ea
So,
\be
a^{-2}(a^2\om_{ab}\dot{)}=\cs\ta\om_{ab}.
\ee
This is a first order differential equation and its solution is
\ba
\label{omegaab} \om_{ab}=\Om_{ab}\frac{1}{\eo a^2}e^{\int\cs\ta dt}=\frac{\Om_{ab}}{\eo a(a-\bt)}, \quad \dot{\Om}_{ab}=0.
\ea

The vector part of shear ($\sv_{ab}$) is obtained from (\ref{sivector}). Let us define a dimensionless and spatial derivative operator $D_a$ as
\bn
D_a = a\eo\sd, \quad D^2 = a^2\eo^2\lp. 
\en
$D_a$ commutes with the derivative along fluid flow lines $u^a\nabla_a$.

Then,
\be
r_a=\nabla^b\om_{ab}=\eo^{-1}a^{-1}D^b\om_{ab}=\frac{R_a}{\eo^2a^2(a-\bt)},
\ee
where $R_a=D^b\Om_{ab}$, $\dot{R}_a=0$.

So,
\ba
\label{svsol} \sv = 2\frac{\eo^{-2}R}{a^2(a-\bt)}\frac{a}{k} = \frac{2R}{\eo q a(a-\bt)} 
\ea


\subsection{Gravitational waves}

The pure tensor parts of $\st_{ab}$, $E^T_{ab}$ and $H^T_{ab}$ are the gravitational waves. The linearized equation for $\st_{ab}$ is obtained from (\ref{sdot})-(\ref{hdot}) by setting $X_{ia}=Z_a=0$, $\om_{ab}=0$,
\ba
\label{trisi} \triangle \st_{ab}+\frac{5}{3}\ta\dot{\st}_{ab}+\frac{1}{6}(\ta^2-9\kappa p)\st_{ab}=0.
\ea
$E^T_{ab}$ and $H^T_{ab}$ are given by,
\ba
E^T_{ab} = -a^{-2}(a^2\st\dot{)}_{ab}, \quad H^T_{ab} = -\mc\st_{ab}.
\ea

Using dimensionless variables, (\ref{trisi}) takes the following form:
\be
\label{stdprime} {\st}''(q,x)+\frac{8x}{\xa}{\st}'(q,x)+\lt\frac{6}{\xa}+q^2\rt\st(q,x)=0.
\ee
The general solution for the $q=0$ mode is
\ba
\st(0,x)=D_1(0)\frac{x(3x^4+10x^2+15)}{3(\xa)^3} \nonumber \\
 +D_2(0)\frac{1}{(\xa)^3}.
\ea

\begin{figure}\label{alpha_vs_x}
\begin{center}
\includegraphics[height=9cm,angle=-90]{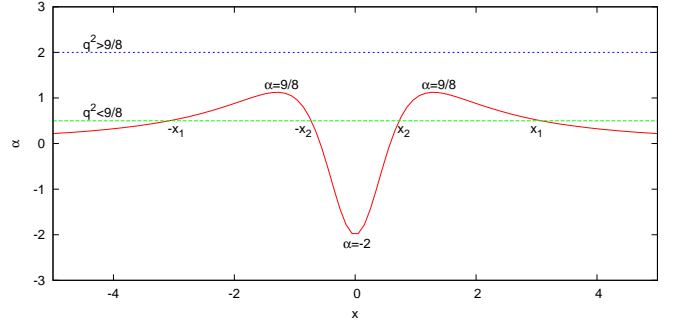}
\caption{Plot of $\alpha$ as a function of $x$. $\alpha$ has a minimum value $\alpha_{\mbox{min}}=-2$ at the bounce and two maxima $\alpha_{\mbox{max}}=\frac{9}{8}$ at $x=\pm\sqrt{\frac{5}{3}}$. For $q^2<\frac{9}{8}$, there are two regions where $q^2-\alpha<0$. But for $q^2>\frac{9}{8}$, $q^2-\alpha$ is always positive.}
\label{alphaplot}
\end{center}
\end{figure}

Using the variable, $f=(1+x^2)\si^T$, Eq. (\ref{stdprime}) becomes,
\be
f''+\Lt q^2-\alpha(x)\Rt f=0, \quad \alpha(x)=2\frac{3x^2-1}{(x^2+1)^2}
\ee
For $x^2\gg\frac{6}{q^2}$, $f$ oscillates with frequency $q$. If $q^2<\frac{9}{8}$, the equation 
\be
q^2-\alpha(x)=0
\ee
has four roots, $\pm x_1(q), \pm x_2(q)$. For $x_2<|x|<x_1$, $q^2-\alpha(x)$ is negative, but $f$ oscillates again for $-x_2<x<x_2$. If however $q^2>\frac{9}{8}$, $q^2-\alpha$ is positive always and $f$ shows oscillatory behavior over the whole range of $x$. The frequency of oscillation is maximum at the point of bounce $x=0$. Graphical representation of $\alpha(x)$ is shown in \figurename{(\ref{alphaplot})}. In any case $f$ and hence $\si^T$ never blow up at bounce or at turning points.

\vspace{5mm}
\textbf{Region A}:
In this region, (\ref{stdprime}) becomes
\be
x^2{\st_A}''+8x{\st_A}'+\lt q^2x^2+6\rt{\st_A}=0.
\ee
The general solution of $\si^T$ in this region is
\be
\st_{A}(q,x)=(qx)^{-7/2}\Lt D_1^A(q)J_{5/2}(qx)+D_2^A(q)Y_{5/2}(qx) \Rt,
\ee
where $J$ and $Y$ are the Bessel function and the Neumann function respectively.

\vspace{5mm}
\textbf{Region B}: 
Using the variable $y=\sqrt{3}x+1$ in region B, Eq. (\ref{stdprime}) is simplified to
\be
\frac{d^2\st_B}{dy^2}-2\frac{d\st_B}{dy}+\lt\frac{3}{2}+\frac{q^2}{3}\rt\st_B=0
\ee

and its general solution is
\be
\label{stsol_B}\st_B(q,y)=e^{y}\Lt D_1^B(q)\cos\lt m_qy\rt+D_2^B(q)\sin\lt m_qy\rt \Rt,
\ee
where,
\be
m_q=\sqrt{\fa+\frac{q^2}{3}}.
\ee

\vspace{5mm}
\textbf{Region C}: 
At the bounce, $(x\rightarrow 0)$, as explained earlier, $\si^T$ oscillates with frequency $\sqrt{2+q^2}$:
\be
\st_C(q,x)=D_1^C(q)\cos\lt\sqrt{2+q^2}x\rt+D_2^C(q)\sin\lt\sqrt{2+q^2}x\rt.
\ee

\vspace{5mm}


\section{Comoving curvature perturbation}
\label{curvature}

The comoving curvature perturbation is defined as \cite{Langlois:2005ii,Langlois:2010vx}
\be
\label{zetaa} \zeta_a = W_a + \frac{X_a}{3\kappa(\mu+p)}, \quad W_a =\sd \log a.
\ee

This variable is related to the comoving curvature perturbation $\zeta$, used in the coordinate based perturbation theory \cite{Mukhanov:1990me}. In particular, since $\zeta_a$ is a spatial gradient of scalar up to first order, we can write
\be
\zeta_a=\sd\zeta^S.
\ee
In \appendixname{~\ref{ordinary}} it has been shown that $\zeta^S$ is equal to $-\zeta$ on the large scale. $\zeta^S$ is conserved on all scales for adiabatic perturbation, whereas $\zeta$ is conserved on the large scale only. However in our model adiabatic modes are present. So the evolution of $\zeta_a$ is determined by the following equation:
\ba
\lu\zeta_a &=& -\frac{\ta}{3\kappa(\mu+p)}\Gamma_a, \\
\label{zedot} \Rightarrow a^{-1}h_a^{~b}(a\zeta_b\dot{)} &=& -\frac{\ta}{3\kappa(\mu+p)}\Gamma_a \nonumber \\
&& \qs -(\si^b_{~a}+\om^b_{~a})\zeta_b,
\ea

$\lu$ being the Lie derivative with respect to $u^a$. Up to first order, using (\ref{Gamma_qx}),
\ba
a^{-1}(a\zeta_q\dot{)} &=& -\frac{\ta}{3\kappa(\mu+p)}\Gamma = -\frac{\dot{a}}{a}\frac{C_1}{27\eo(a-\bt)^2}.
\ea 
Integrating,
\ba
\label{zetasol} \zeta_q=\frac{1}{27\eo}\frac{1}{a}\lt\frac{C_1}{a-\bt}+\tilde{C}_2\rt.
\ea
So, besides the nonadiabatic constant mode of $\zeta^S\sim -\frac{a}{k}\zeta_q$ there is an adiabatic mode which diverges at the turning point.


\section{Validity of linear treatment at the turning point}
\label{compare}

The speed of sound and different perturbation variables become infinite at the turning point, not at the bounce. Existence of these growing modes raised doubts on the validity of linear perturbation theory. In the coordinate based perturbation theory, linear perturbation treatment is justified if the perturbations remain small compared with background quantities. However in covariant perturbation theory, background values of all gauge invariant variables are zero. So in this case we demand that higher order terms in the perturbation equation must be small compared with the first order term. Let us consider the equations for scalar perturbation (\ref{xdot}) and (\ref{zdot}). We have defined some linearity parameters $\varepsilon_1-\varepsilon_7$ and $\tilde{\varepsilon}_3-\tilde{\varepsilon}_7$ as the ratio of nonlinear to the linear terms in these equations in \cite{Kumar:2012tr}. The linear perturbation theory for the scalar perturbations is valid, if the following conditions are satisfied throughout the regime under consideration:
\ba
 & & \mathbf{(1)} \; \varepsilon_1, \varepsilon_2 \ll 1, \nonumber \\
 & & \mathbf{(2)} \; \varepsilon_3 , \varepsilon_4, \varepsilon_5, \varepsilon_6, \varepsilon_7 \ll 1 ~\mbox{and}/\mbox{or}~ \tilde{\varepsilon}_3, \tilde{\varepsilon}_4, \tilde{\varepsilon}_5, \tilde{\varepsilon}_6, \tilde{\varepsilon}_7 \ll 1. \nonumber
\ea

Using Eq. (\ref{allscalar}) and the solutions (\ref{wBqy}), (\ref{zBqy}), (\ref{omegaab}), (\ref{srsol}), (\ref{svsol}) and (\ref{stsol_B}) of the perturbation equations at region B, the dominant mode of different variables that appear in (\ref{xdot}) and (\ref{zdot}) can be written as
\ba
& & X_a = \frac{1}{\eo^3\bt^4}\lt\Upsilon_a+\Xi_a\rt, \quad 
Z_a = \frac{4}{27\eo^2\bt^3}\frac{D^2\Xi_a}{y}, \nonumber\\
& & \nu_a = \frac{4}{27\eo\bt^2}\frac{\Xi_a}{y^2}, \quad 
A_a = \frac{4}{27\eo^3\bt^4}\frac{D^2\Xi_a}{y^2}, \nonumber\\
& & \si_{ab} = \frac{4}{27\eo\bt^2}\frac{D_{<a}\Xi_{b>}+D_{<a}\Lambda_{b>}}{y}, \quad
\om_{ab} = -\frac{2}{\eo\bt^2}\frac{\Om_{ab}}{y}, \nonumber\\
& & \mathcal{R} = \frac{4}{27\eo^2\bt^3} \lt D^a\Xi_a + \frac{4}{27\bt}|D_{<a}\Xi_{b>}+D_{<a}\Lambda_{b>}|^2 -\frac{4}{\bt}|\Om_{ab}|^2 \rt \frac{1}{y^2}, \nonumber\\
& & 2\sd\si^2 = \frac{4}{27\eo^3\bt^5}\frac{D_a\lt|D_{<c}\Xi_{d>}+D_{<c}\Lambda_{d>}|^2\rt}{y^2}, \nonumber\\
& & \label{XtoZ@B} 2\sd\om^2 = \frac{4}{\eo^3\bt^5}\frac{D_a\lt|\Om_{cd}|^2\rt}{y^2},
\ea

where
\ba
& \Upsilon_a = \sum_k C_1(q)Q^{(0)}_a , \quad \Xi_a = -\fa\sum_k C^B_3(q)Q^{(0)}_a, \nonumber \\
& \Lambda_a=27\sum_k\frac{R(q)}{q^2}Q^{(1)}_a.
\ea

At the turning point, $y\sim 0$, (\ref{muplusp}) reduces to 
\be
\kappa(\mu+p)=-\frac{\kappa M_1}{2\bt^3}y= -\frac{9}{2}\eo^{-2}\bt^{-2}y.
\ee

Then the linearity parameters for (\ref{xdot}) and (\ref{zdot}) are found to be
\begin{widetext}
\ba
& & \varepsilon_1 = \frac{\lb\om^b_{~a}X_b\rb}{\lb\kappa(\mu+p)Z_a\rb} 
                  = \frac{3}{\bt}\frac{\lb\Om^b_{~a}\lt\Upsilon_b+\Xi_b\rt\rb}{\lb D^2\Xi_a \rb} y^{-1} , \quad\quad
 \varepsilon_2 = \frac{\lb\si^b_{~a}X_b\rb}{\lb\kappa(\mu+p)Z_a\rb} 
                  = \frac{2}{9\bt}\frac{\lb h^{bc}\lt D_{<a}\Xi_{b>}+D_{<a}\Lambda_{b>}\rt\lt\Upsilon_c+\Xi_c\rt\rb}{\lb D^2\Xi_a \rb} y^{-1} , ~\nonumber \\
& & \varepsilon_3 = \frac{\lb\mathcal{R}\nu_a\rb}{\lb\fa X_a\rb} 
                  = \lt\fc\rt^5\frac{1}{3\bt}\frac{\lb D^c\Xi_c + \frac{4}{27\bt}|D_{<c}\Xi_{d>}+D_{<c}\Lambda_{d>}|^2 -\frac{4}{\bt}|\Om_{cd}|^2 \rb\lb\Xi_a\rb}{\lb\Upsilon_a+\Xi_a\rb} y^{-4} , \nonumber \\,
& & \varepsilon_4 = \frac{\lb 2h_a^{~b}\nabla_b\om^2\rb}{\lb\fa X_a\rb} 
                  = \frac{8}{\bt}\frac{\lb D_a\lt|\Om_{cd}|^2\rt\rb}{\lb\Upsilon_a+\Xi_a\rb}y^{-2} ,  \quad\quad
 \varepsilon_5 = \frac{\lb 2h_a^{~b}\nabla_b\si^2\rb}{\lb\fa X_a\rb}
                  = \lt\fc\rt^3\frac{1}{\bt}\frac{\lb D_a\lt|D_{<c}\Xi_{d>}+D_{<c}\Lambda_{d>}|^2\rt\rb}{\lb\Upsilon_a+\Xi_a\rb}  y^{-2} ,  \nonumber \\
& & \varepsilon_6 = \frac{\lb\om^b_{~a}Z_b\rb}{\lb\fa X_a\rb}
                  = \lt\fc\rt^3\frac{2}{\bt}\frac{\lb\Om^b_{~a}D^2\Xi_b\rb}{\lb\Upsilon_a+\Xi_a\rb}y^{-2},  \quad\quad
 \varepsilon_7 = \frac{\lb\si^b_{~a}Z_b\rb}{\lb\fa X_a\rb} 
                  = \lt\fc\rt^3\frac{1}{3\bt}\frac{\lb h^{bc}\lt D_{<a}\Xi_{b>}+D_{<a}\Lambda_{b>}\rt D^2\Xi_b\rb}{\lb\Upsilon_a+\Xi_a\rb}  y^{-2} .
\ea

Other sets of parameters $\tilde{\varepsilon}_3$-$\tilde{\varepsilon}_7$ are related to the $\varepsilon_3$-$\varepsilon_7$ via

\be
\tilde{\varepsilon}_I = \fa \frac{|X_a|}{|A_a|}\varepsilon_I = \lt\frac{3}{2}\rt^3\frac{\lb\Upsilon_a+\Xi_a\rb}{\lb D^2\Xi_a\rb}y^2\varepsilon_I, \quad \mbox{for I=3 ~\text{to}~ 7}.
\ee

\ba
& & \tilde{\varepsilon}_3 = \frac{4}{27\bt}\frac{\lb D^c\Xi_c + \frac{4}{27\bt}|D_{<c}\Xi_{d>}+D_{<c}\Lambda_{d>}|^2                -\frac{4}{\bt}|\Om_{cd}|^2 \rb\lb\Xi_a\rb}{\lb D^2\Xi_a\rb} y^{-2} , \qs\quad \nonumber \\
& & \tilde{\varepsilon}_4 = \frac{27}{\bt}\frac{\lb D_a\lt|\Om_{cd}|^2\rt\rb}{\lb D^2\Xi_a\rb} ,  \qs
 \tilde{\varepsilon}_5 = \frac{1}{\bt}\frac{\lb D_a\lt|D_{<c}\Xi_{d>}+D_{<c}\Lambda_{d>}|^2\rt\rb}{\lb D^2\Xi_a\rb},  \nonumber \\
& & \tilde{\varepsilon}_6 = \frac{2}{\bt}\frac{\lb\Om^b_{~a}D^2\Xi_b\rb}{\lb D^2\Xi_a\rb},  \qs
 \tilde{\varepsilon}_7 = \frac{1}{3\bt}\frac{\lb h^{bc}\lt D_{<a}\Xi_{b>}+D_{<a}\Lambda_{b>}\rt D^2\Xi_b\rb}{\lb D^2\Xi_a\rb}.
\ea

\end{widetext}

The $\varepsilon_1$ and $\varepsilon_2$ diverge at the turning point as $y\rightarrow 0$. So the condition (1) is not satisfied at the turning point. Although $\tilde{\varepsilon}_4-\tilde{\varepsilon}_7$ remain finite at the turning point, $\tilde{\varepsilon}_3$ diverges. So the condition (2) is also not satisfied.


\section{Matching condition}
\label{matching}

We have seen that even for this simple model analytical expressions for the perturbation variables throughout the bounce are not available. One can obtain the solutions by numerical integration but to have a good understanding on the result one needs some analytical methods. Such methods involve matching of the variables across the transition surfaces. In the non-bouncing cases it is well known that the spatial metric on the hypersurface and the extrinsic curvature must be continuous across the boundary separating the two regions \cite{Deruelle:1995kd}. However, for the bouncing models one should find the appropriate variables, which should be matched to get a correct spectrum. In a nonsingular bouncing background the spatial curvature perturbation $\zw$ is found to be the appropriate variable (rather than the Bardeen potential $\Phi$) which is to be matched in order to get good agreement with the numerical results \cite{Gasperini:2004ss}. In \appendixname{~\ref{ordinary}} we have shown that $\zw$ and $\Phi$ are related to $\vw_a$ and $X_a$ respectively. We now investigate whether matching of these variables will lead to the correct spectrum after the bounce. 

Considering only scalar variables, we have
\ba
\label{xvw} X_i &=& \frac{2}{a^2}\pa_i\vn\Phi, \quad \vw_i = \frac{2}{a^2}\pa_i\vn\zw.
\ea

We consider the perturbation modes that exit the horizon in deep matter dominated era ($|x|\gg 1$). If $x=-\ex$ is the value of $x$ at the horizon exit, then
\be
q=|\mh_{\mbox{exit}}| \quad \Rightarrow \quad \ex=\frac{2}{q}.
\ee
Since $\ex\gg 1$, $q$ must be much less than order unity. Expanding $\vw_a$ and $\zw$ in Fourier modes and considering only scalar modes,
\be
\vw_a = \sum_k \eo^{-3}a^{-a}\mv Q^{(0)}_a, \quad \zw = \sum_k \zw_q Q^{(0)} 
\ee

Then (\ref{xvw}) leads to
\be
\label{mvzw} \mv \approx 2q^3\zw_qa
\ee

$\mv$ can be written in terms of $\mx$ and $\mz$ as,

\be
\label{mvmxmz} \mv=\lt1+\frac{2q^2a^2}{27\bt(a-\bt)}\rt\mx-2\mh\mz
\ee

The Mukhanov-Sasaki variable is defined as $v=\zw_qz$, where $z=3a\ta^{-1}\sqrt{\kappa(\mu+p)}$. In our model,
\be
z=\sqrt{3}a\sqrt{\frac{a-\bt}{a-\epsilon}},
\ee
\be
\mv \approx \frac{2}{\sqrt{3}}q^3\sqrt{\frac{a-\epsilon}{a-\bt}}v.
\ee

The initial values of $v$ and its derivative are given by the quantum vacuum initial condition at the time of horizon exit:

\be
v\sim \sqrt{\frac{1}{2q}}, \quad v' \sim i\sqrt{\frac{q}{2}} 
\ee

In this region, $a \gg \bt, \epsilon$. So, $\mv\approx\frac{2}{\sqrt{3}}q^3v$ and the initial conditions on $\mv$ are obtained as
\be
\label{mvini}\mv\sim\sqrt{\fc q^5}, \quad \mv'\sim i\sqrt{\fc q^7}.
\ee

Now in region A of contracting phase, 
\ba
\label{mxminus} & & \mx^{(-)}=\frac{2q^2}{27}C^{(-)}_1+\frac{C^{A(-)}_2}{x^3}+C^{A(-)}_3x^2, \\
& & \mv^{(-)} = \fd \frac{C^{(-)}_1}{x^2}\lt1+\frac{q^2x^2}{36}\rt\lt1+\frac{q^2x^2}{18}\rt + \frac{q^2}{36}\frac{C^{A(-)}_2}{x} \nonumber \\
\label{mvminus} & & \ql + \frac{5}{3}C^{A(-)}_3x^2\lt1+\frac{q^2x^2}{60}\rt.
\ea

In the expanding phase perturbations have similar evolution but with different constants,
\ba
\label{mxplus} && \mx^{(+)}=\frac{2q^2}{27}C^{(+)}_1+\frac{C^{A(+)}_2}{x^3}+C^{A(+)}_3x^2, \\
&& \mv^{(+)} = \fd \frac{C^{(+)}_1}{x^2}\lt1+\frac{q^2x^2}{36}\rt\lt1+\frac{q^2x^2}{18}\rt + \frac{q^2}{36}\frac{C^{A(+)}_2}{x} \nonumber \\
\label{mvplus} && \ql + \frac{5}{3}C^{A(+)}_3x^2\lt1+\frac{q^2x^2}{60}\rt.
\ea

The relation between the constants are obtained by proper matching of the variables in the boundary of the bouncing phase. We want to study such matching conditions on the surfaces $x=\pm 1$, which are the boundary of week energy condition ($\mu+3p\ge0$) violated region. First we deduce the spectrum of perturbations using two matching conditions, namely the continuity of $\mv$ and $\mx$ across the transitions surface and then calculate the same spectrum from numerical computation.   

Since the entropy perturbation is obtained for all values of $a$, we get a matching condition, 
\be
C^{(+)}_1=C^{(-)}_1
\ee

Matching $\mv$ and $\mv'$ on these surfaces, we get
\ba
\frac{q^2}{12}C^{A(+)}_2 &=& -\frac{16}{3}C^{(-)}_1 - \frac{q^2}{36}C^{A(-)}_2 + \frac{20}{3}C^{A(-)}_3 \nonumber \\
\label{matchV} 5C^{A(+)}_3 &=& \frac{16}{3}C^{(-)}_1 - \frac{q^2}{18}C^{A(-)}_2 - \frac{5}{3}C^{A(-)}_3
\ea

To know the correct spectrum of perturbation, we need the initial conditions on non adiabatic perturbations. For simplicity, let us assume $\Gamma_a=0$, which implies, by (\ref{Gamma_qx}), $C^{(-)}_1=0$. Then the initial conditions (\ref{mvini}) give
\ba
\label{Cini} C^{A(-)}_2\approx(i-1)8\sqrt{\fc}q^{-1/2}, \nonumber \\ C^{A(-)}_3\approx\frac{2i-1}{8}\sqrt{\fc}q^{9/2}.
\ea

Then (\ref{matchV}) leads to
\ba
C^{(-)}_1=0, \quad C^{A(+)}_2\approx(1-i)\frac{8}{3}\sqrt{\fc}q^{-1/2}, \\ C^{A(+)}_3\approx(1-i)\frac{4}{45}\sqrt{\fc}q^{3/2}.
\ea

Using this constants in (\ref{mvplus}), we get

\be
\mv^{(+)} \approx (1-i)\frac{2}{27}\sqrt{\fc}q^{3/2}\lt\frac{1}{x}+2x^2\lt1+\frac{q^2x^2}{60}\rt\rt.
\ee

In the deep matter dominated phase,
\be
\label{spectra_vv}\left|\mv^{(+)}\right|^2 \approx q^3\left|1+\frac{q^2x^2}{60}\right|^2.
\ee

Using (\ref{mvzw}) the spectrum of $\zw_q$ is found to be

\be
P_{\zeta} \approx q^3\left| \zw_q \right|^2 \approx \left|1+\frac{q^2x^2}{60}\right|^2.
\ee

So the power spectrum of $\zw$, obtained from this matching condition is nearly scale invariant, provided $q^2x^2<60$, which is satisfied even after the horizon reentry ($qx=2$). Using this matching we can also calculate the spectrum of $\mx$. From (\ref{mxplus}),

\ba
\mx^{(+)}= -\fd\sqrt{\fc}(i-1) \lt 2q^{-1/2}x^{-3} + \frac{q^{3/2}x^2}{15} \rt.
\ea

So in the deep matter dominated era of the expanding phase, 
\be
\label{spectra_xv} \left|\mx^{(+)}\right|^2 \approx q^3.
\ee

Now we use a different matching condition, i.e. matching of $\mx$. That leads to, using (\ref{Cini}),
\ba
C^{(+)}_1=0, \quad C^{A(+)}_2 \approx (1-i)\frac{8}{5}\sqrt{\fc}q^{-1/2}, \\ C^{A(+)}_3 \approx -(1-i)\frac{16}{15}\sqrt{\fc}q^{-1/2}.
\ea

The spectra of $\mv$ and $\mx$ are found to be

\ba
\label{spectra_x}\left|\mv^{(+)}\right|^2 \approx q^{-1}\left|1+\frac{q^2x^2}{60}\right|^2, \quad \left|\mx^{(+)}\right|^2 \approx q^{-1}.
\ea

We will find that the numerical results agree with (\ref{spectra_vv}) and (\ref{spectra_xv}), not with (\ref{spectra_x}).


\begin{figure}
\begin{center}
\includegraphics[height=9cm,angle=-90]{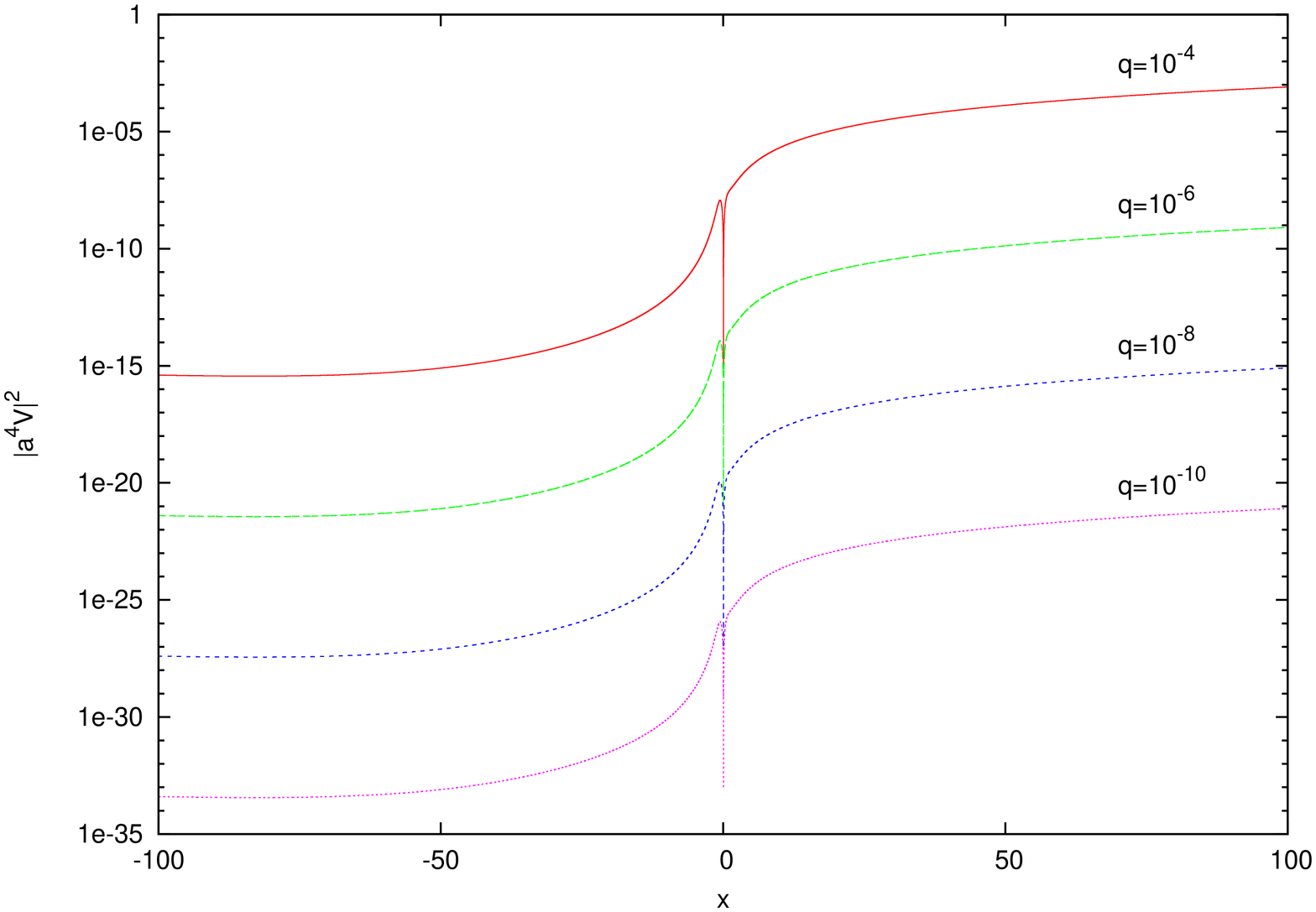}
\caption{Time evolution of $\mv$ with $x$ with different values of wave number $q$.}
\label{spectrum_V}
\end{center}
\end{figure}

\begin{figure}
\begin{center}
\includegraphics[height=9cm,angle=-90]{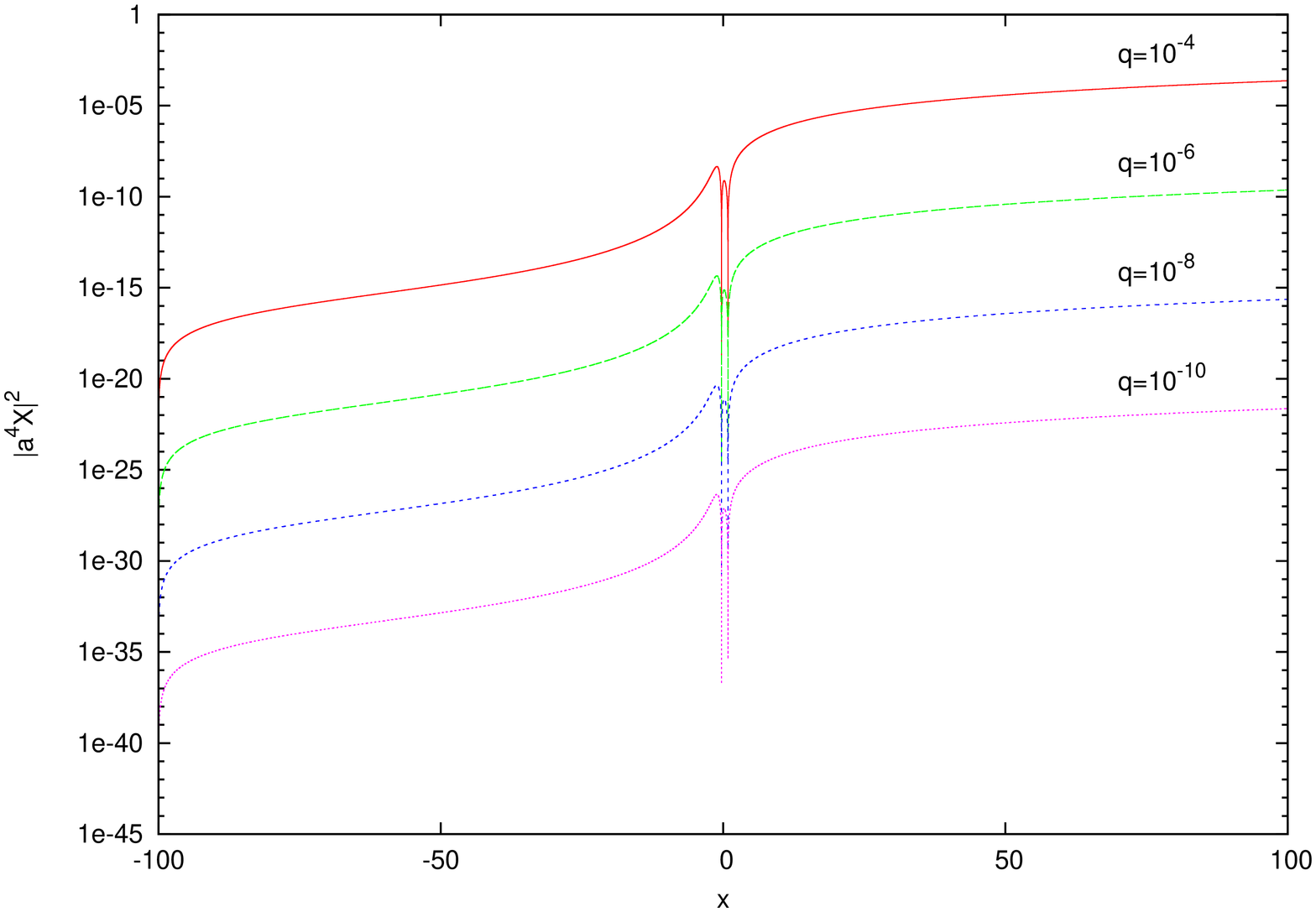}
\caption{Time evolution of $\mx$ with $x$ with different values of wave number $q$.}
\label{spectrum_X}
\end{center}
\end{figure}

\section{Numerical Analysis}
\label{numerical}

We solve the coupled set of differential equations (\ref{x1pri})-(\ref{zpri}) by the Runge-Kutta method. The initial conditions are chosen as follows. The perturbations exit the horizon at $x=-\ex$ in the matter dominated era. At a later time $x=-x_0$, but still within the matter dominated era, $\mv$ and $\mv'$ are given by

\ba
\mv(-x_0)\approx\frac{\ex}{x_0}\mv(-\ex)=\frac{2}{x_0}\sqrt{\fc q^3} \nonumber\\
\label{mv0}\mv'(-x_0)\approx\lt\frac{\ex}{x_0}\rt^2\mv'(-\ex)=i\frac{2}{x_0}\sqrt{\fc q^3},
\ea
where we have used the initial conditions (\ref{mvini}).

\begin{figure}
\begin{center}
\includegraphics[height=9cm,angle=-90]{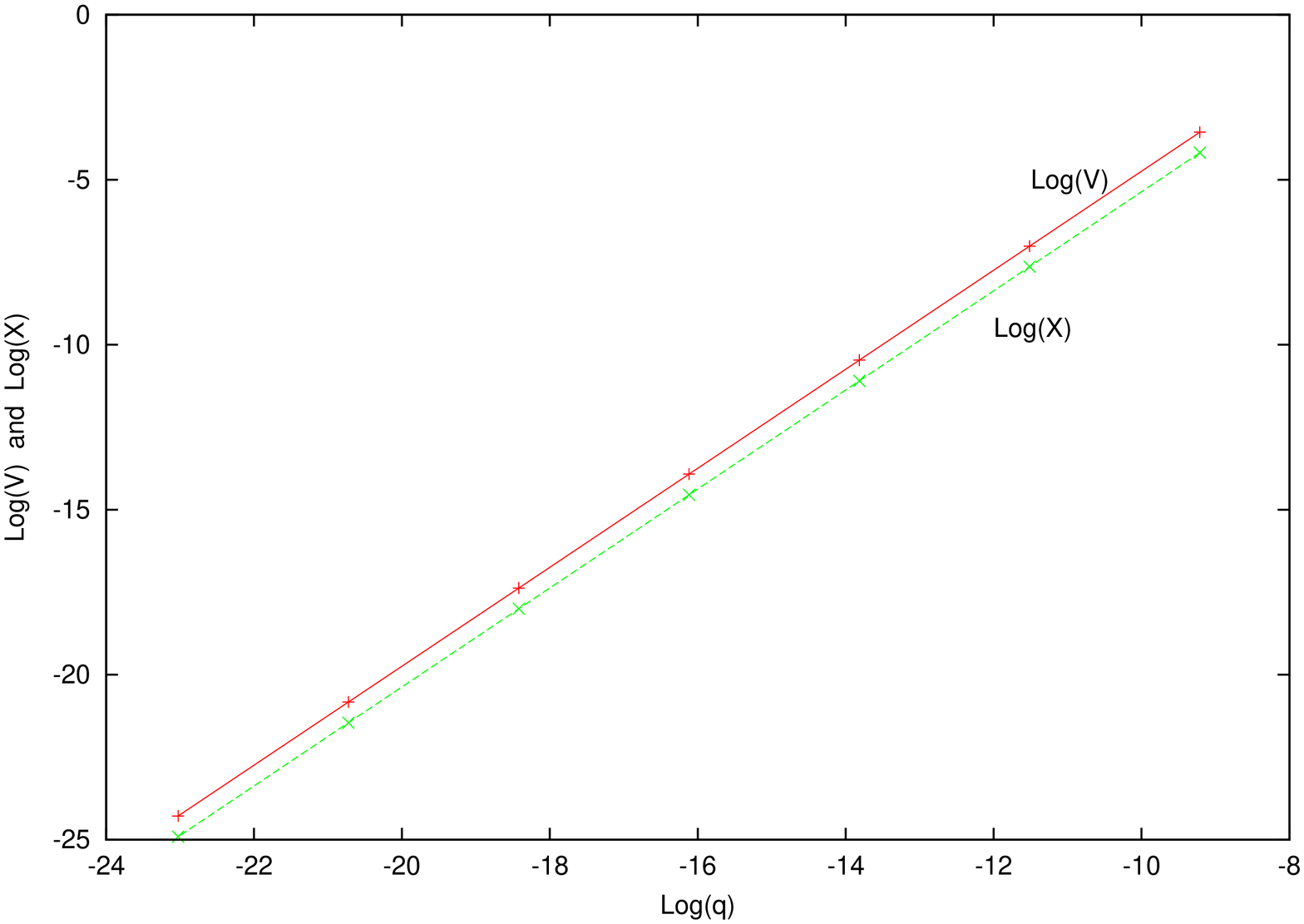}
\caption{Spectral distribution of $\mx$ and $\mv$ at a fixed time $x=100$.}
\label{spectrum_XV}
\end{center}
\end{figure}

Now since $C_1=0$, 
\ba
\mx_2=\frac{\bt}{a}\mx_1, \quad \mx=\frac{a-\bt}{a}\mx_1.
\ea

From (\ref{mvmxmz}) and using (\ref{x1pri})-(\ref{zpri}) we get

\bn
\mv=\mathbf{A}\mx_1+\mathbf{B}\mz, \\ 
\mv'=\mathbf{C}\mx_1+\mathbf{D}\mz,
\en

where,
\bn 
\mathbf{A}=\frac{a-\bt}{a}+\frac{2q^2}{27}\frac{a}{\bt}, \quad \mathbf{B}=-2\mh, \\
\mathbf{C}=\frac{2q^2}{27}\frac{a}{\bt}, \quad \mathbf{D}= -6\frac{\bt(a-\epsilon)}{a^2}-\fc q^2
\en

So,
\ba
\mx_1=\frac{\mathbf{D}\mv-\mathbf{B}\mv'}{\mathbf{A}\mathbf{D}-\mathbf{B}\mathbf{C}}, \quad
\label{mxmzabcd} \mz=-\frac{\mathbf{C}\mv-\mathbf{A}\mv'}{\mathbf{A}\mathbf{D}-\mathbf{B}\mathbf{C}}.
\ea

Substituting (\ref{mv0}) in (\ref{mxmzabcd}) we get the values of $\mx_1$, $\mx_2$, $\mz$ at $x=-x_0$. We take $x_0=100$. The results of numerical computation are shown in \figurename{(\ref{spectrum_V}-\ref{spectrum_XV})}. In \figurename{(\ref{spectrum_V})} and \figurename{(\ref{spectrum_X})} the time evolution of $\mv$ and $\mx$ is shown for wave numbers $q=10^{-4},10^{-6},10^{-8},10^{-10}$. It is seen that the spectrum of both variables behaves as $q^{3/2}$ in agreement with (\ref{spectra_vv}) and (\ref{spectra_xv}). We have also plotted $\log|\mx|$ and $\log|\mv|$ as a function of $\log|q|$ in \figurename{(\ref{spectrum_XV})} at a time $x=100$ in the expanding phase when all modes are outside the horizon. This gives
\ba
\frac{\de\log|\mv|}{\de\log|q|}=\frac{\de\log|\mx|}{\de\log|q|}=1.5.
\ea

\begin{figure}
\begin{center}
\includegraphics[height=9cm,angle=-90]{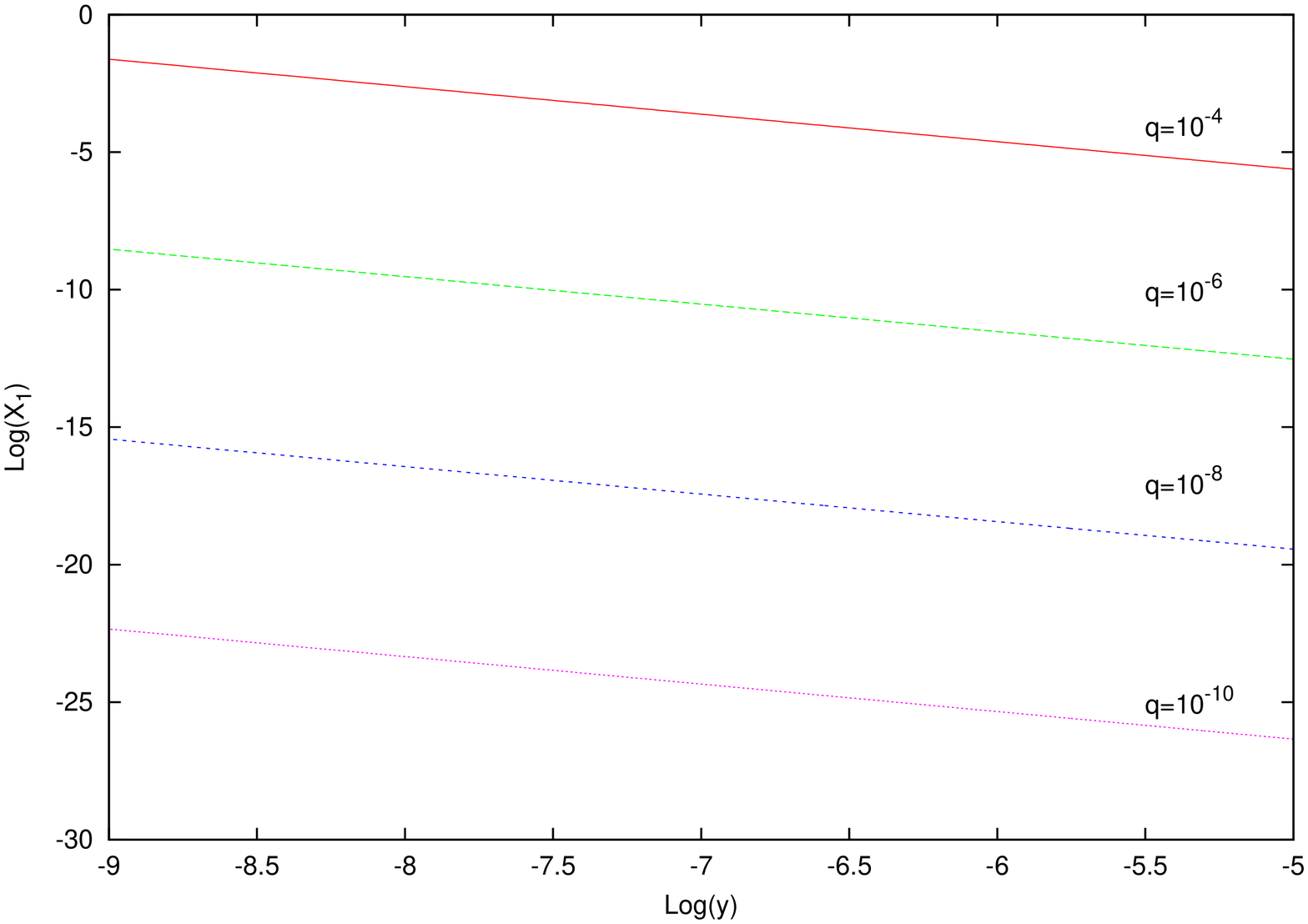}
\includegraphics[height=9cm,angle=-90]{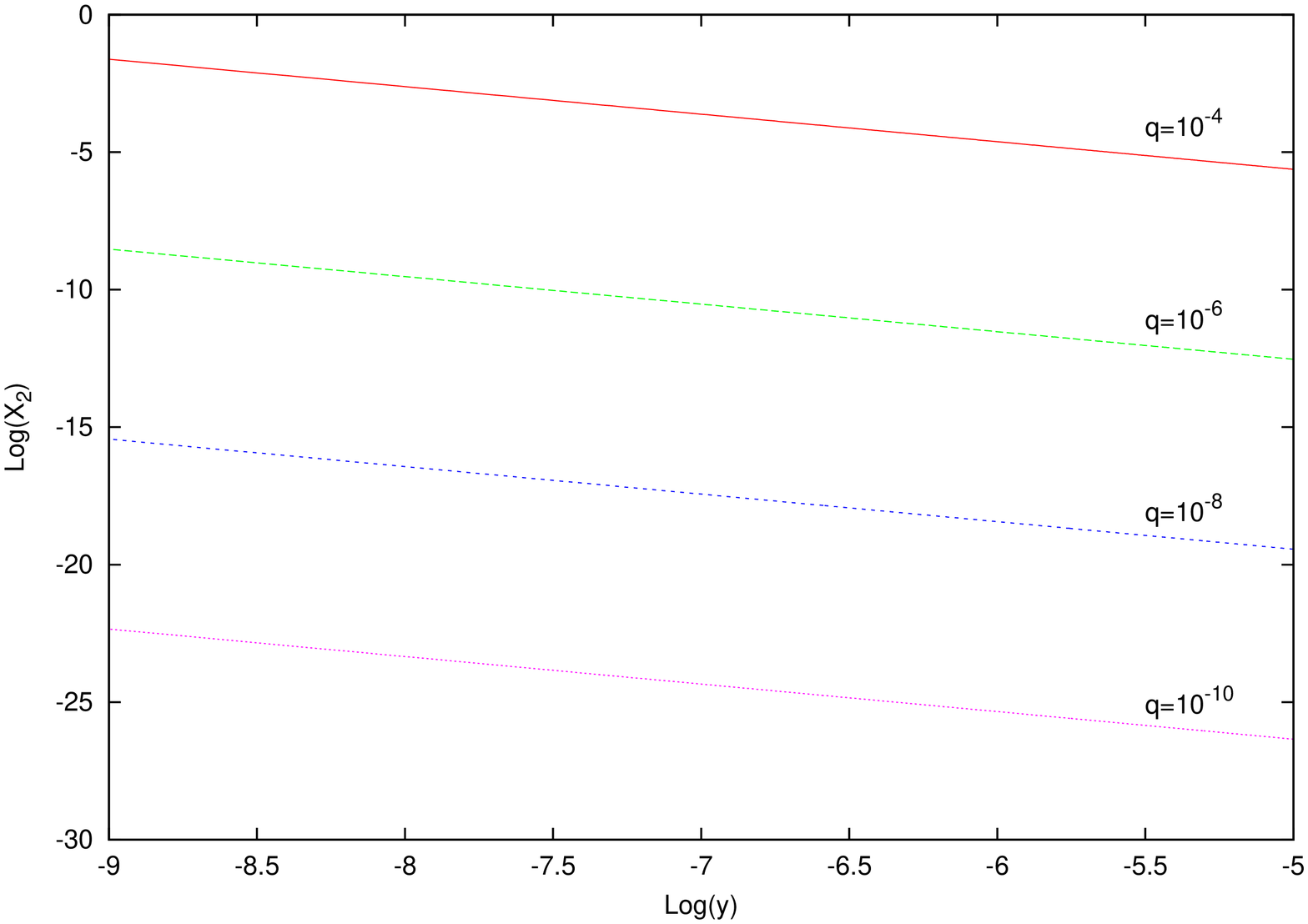}\\
\includegraphics[height=9cm,angle=-90]{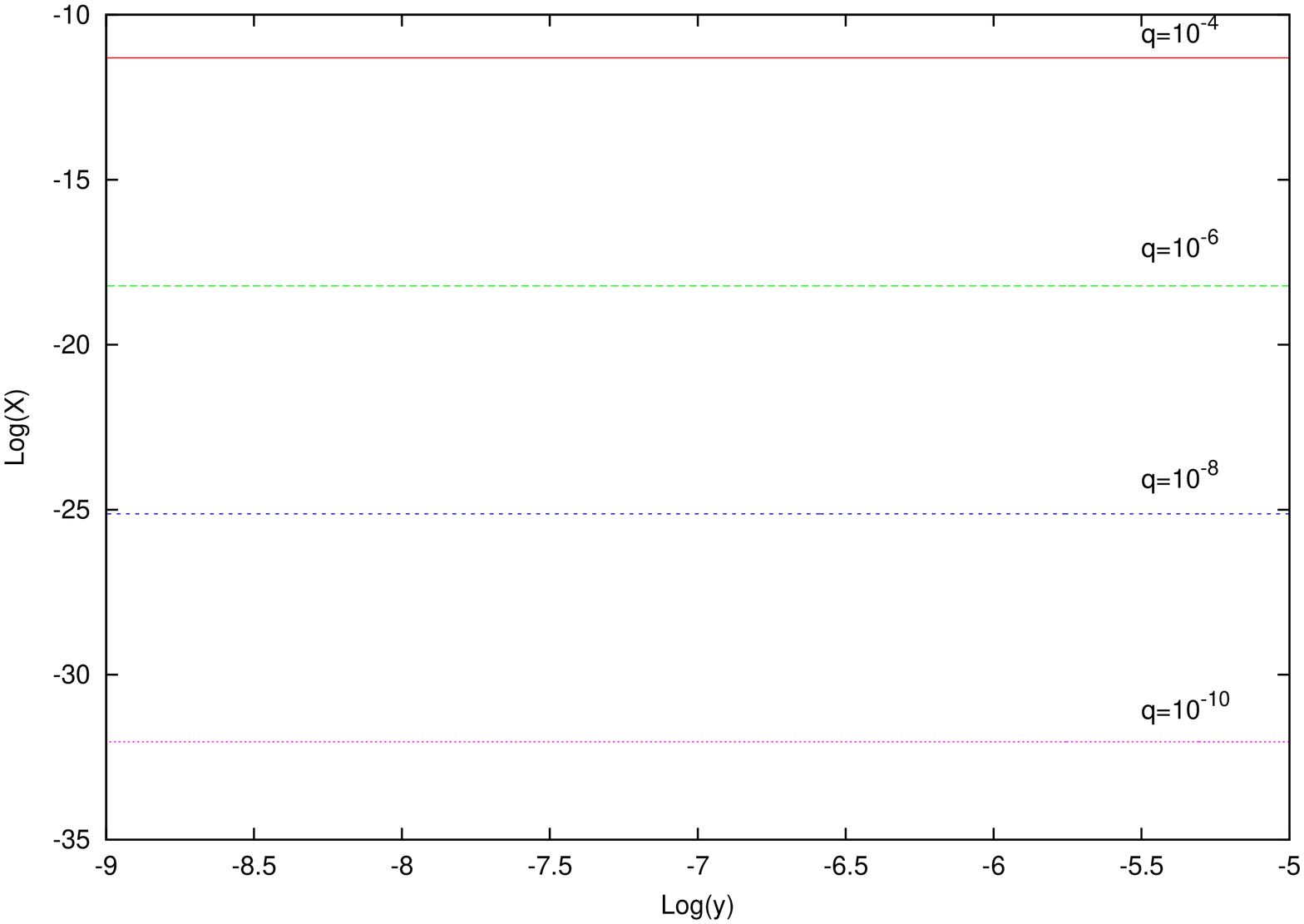}
\includegraphics[height=9cm,angle=-90]{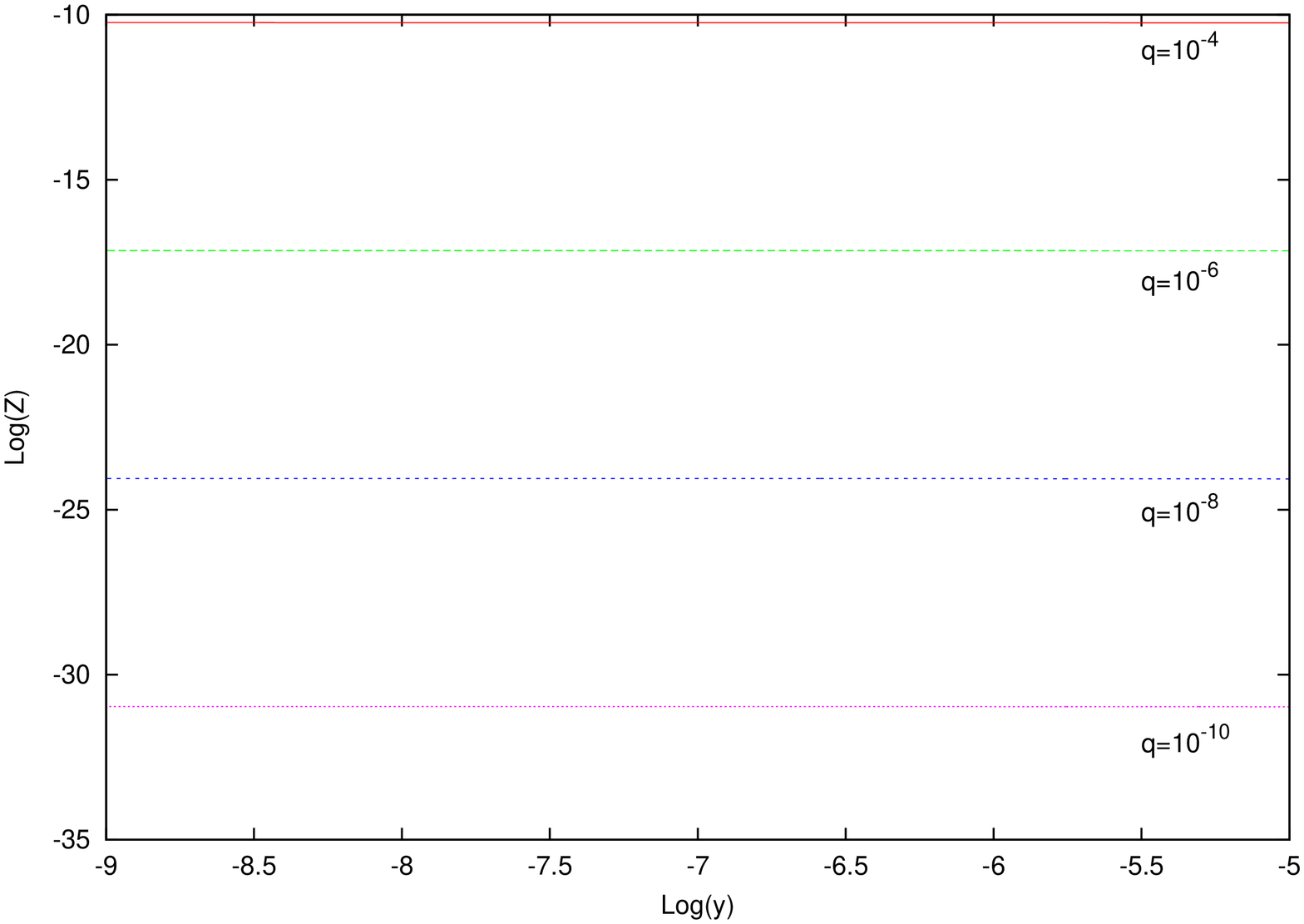}
\caption{Evolution of perturbations near the turning point for different wave numbers.}
\label{XZ@B}
\end{center}
\end{figure}

\begin{figure}
\begin{center}
\includegraphics[height=9cm,angle=-90]{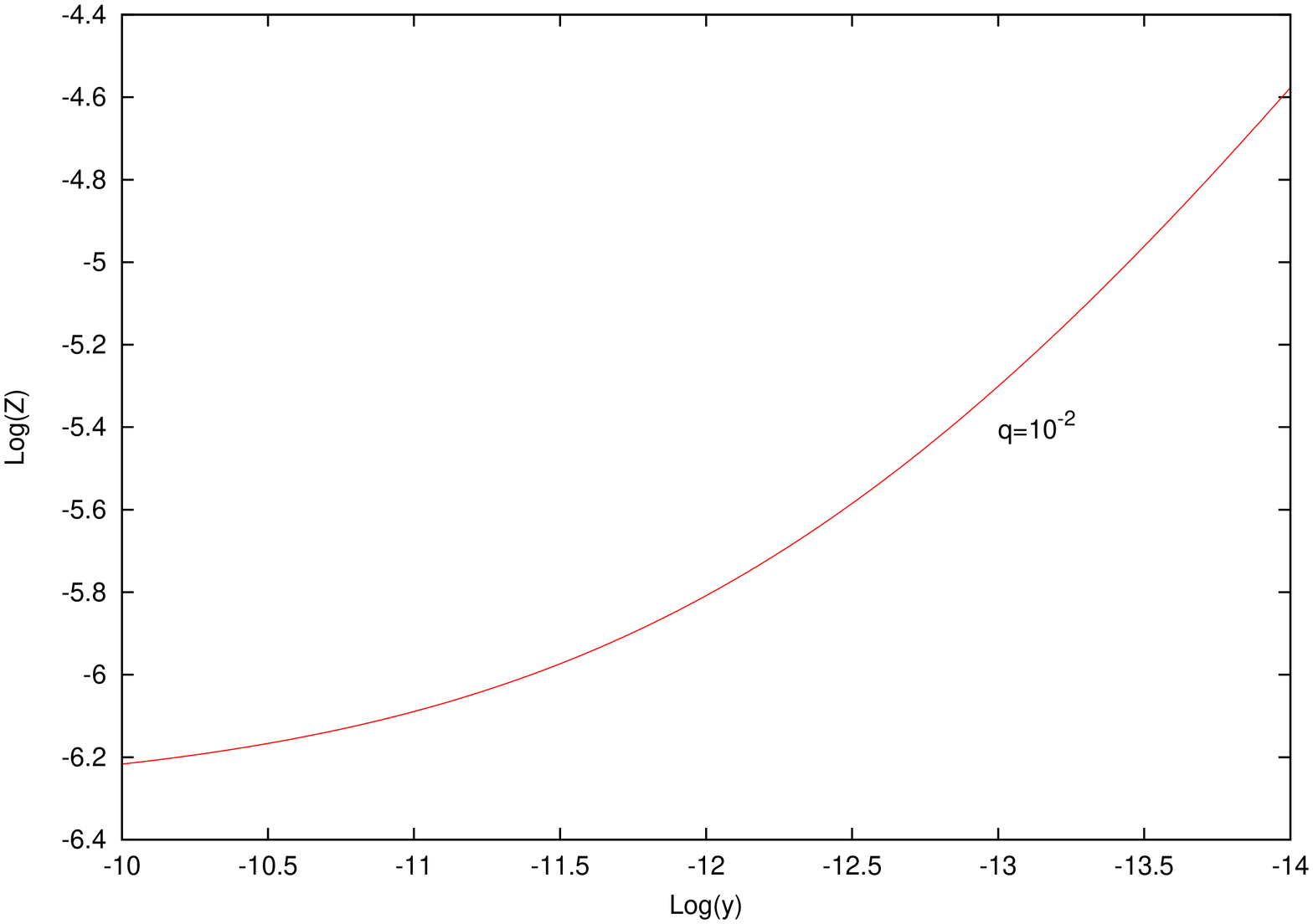}
\caption{Evolution of $Z$ very close to the turning point for $q=0.01$.}
\label{Z@B}
\end{center}
\end{figure}

We have also plotted the behavior of perturbations in region B in \figurename{(\ref{XZ@B})}. It is observed that $X_1$ and $X_2$ grow as $y^{-1}$ near turning point, but $X$ and $Z$ remain constant. However according to (\ref{zBqy}), the growing mode of $Z$ starts to dominate at nearer to the turning point for smaller frequencies. It is evident from \figurename{(\ref{Z@B})} that $Z$ also grows as $y^{-1}$ very close to the turning point.

Hence, the numerical analysis with our special initial conditions support our analytical results (\ref{x1Bqy})-(\ref{zBqy}). Since the growth rates of scalar variables in (\ref{XtoZ@B}) are derived from (\ref{x1Bqy})-(\ref{zBqy}), the results in Sec. \ref{compare} involving scalar variables are still valid.


\section{Conclusion}

We have studied the evolution of cosmological perturbations through a toy model of nonsingular and bouncing universe using the techniques of covariant perturbation theory. The matter sector is a two component perfect fluid. The dustlike normal fluid drives the contraction and expansion and the radiationlike fluid having negative energy density drives the bounce. 

Evolution of vector perturbations $\om_a$ and $r_a$ is rather simple. But the analytic solutions for scalar and tensor perturbations in the entire range of time are obtained only for zero wave number mode. For $q\ne 0$ the equations are simplified to get analytic solutions in three different regions, namely long before bounce, at the turning point and at the bounce. The scalar perturbations are smooth across the bounce but diverge at the turning point. The shear $\si_{ab}$ is decomposed into scalar, vector and pure tensor parts. The gravitational wave, i.e. pure tensor part of shear shows oscillating behavior both at the bounce and at the turning point. At the turning point, scalar and vector parts dominate over the gravitational wave. The comoving curvature perturbation $\zeta^S$ has a nonadiabatic growing mode at the turning point, besides its adiabatic constant mode.  

The growth rates of the linearity parameters are computed at the turning point. It is observed that many of these parameters diverge. So the perturbations cease to be linear at the turning point even in this simple nonsingular bouncing model. 

The perturbation variables are defined here in terms of the velocity $u^a$ of the comoving observers in physical spacetime. This choice is not unique. In order that the perturbations are gauge independent, the variables must vanish in the background spacetime, which means that the world lines of the observers in the physical spacetime must not differ too much from that of the comoving observers in the background spacetime in the following precise sense: One can choose any arbitrary family of observers having velocity $\tilde{u}^a$ (for example, observers whose velocity is normal to the constant energy density hypersurface) such that $\tilde{u}^a-u^a$ vanish in the background spacetime. Let $\tilde{X}_a$, $\tilde{Y}_a$, $\tilde{Z}_a$ etc. be the perturbations, covariantly defined in terms of $\tilde{u}^a$. Then these new variables can be written in terms of the old ones $X_a$, $Y_a$, $Z_a$ etc. Since the evolution equations of the new variables are different from those of the old variables, nonlinearity may not appear in their evolution. So although our results are completely independent of the choice of gauge, they are tied to some choice of observers.

We have studied the matching condition for scalar variables. It has been shown that the spectrum of perturbations after the bounce can be obtained by employing the sound matching condition. Despite the divergence at the turning point and the growth of amplitude, the scale invariant spectrum of the perturbations is preserved after the bounce. Our numerical analysis shows that the variable $\mv$ should be matched across the transition surface to get the correct spectra, while matching $\mx$ will lead to a wrong spectra. Since $\mv$ and $\mx$ are related to spatial curvature perturbation ($\zw$) and the Bardeen potential ($\Phi$), these results coincide with the ones obtained in \cite{Gasperini:2004ss}. 

However, one may ask whether this spectrum is disrupted by the appearance of nonlinearity at the turning point. The $y^{-n}$ dependence of linearity parameters implies that the nonlinearity effect may last only for a very short interval of time. Moreover, the interval may be shorter for larger wavelengths, as indicated by Eq. (\ref{zBqy}) and the numerical analysis. To address the question of whether the temporary nonlinearity can alter the future evolution of perturbations substantially, one requires to perform a full nonlinear analysis as has been performed in \cite{Xue:2013bva} for adiabatic perturbations.

\section*{Acknowledgments}
I would like to thank Amit Ghosh for discussions. Financial support is provided by the Council of Scientific and Industrial Research (CSIR), Government of India.


\appendix
\section{Spatial Harmonics}
\label{harmonics}

The tensor eigenfunctions (harmonics) of the spatial Laplacian $\lp$ listed below, are solutions of the tensor Helmholtz equation:
\begin{equation}
 \lp Q_{ab\cdots} + \frac{k^2}{a^2}Q_{ab\cdots}=0,
\end{equation}
where
\be
\lp Q_{ab\cdots}= h^{pq}h_{a}^{~a_1}h_{b}^{~b_1}\cdots\nabla_p\lt h_q^{~s}h_{a_1}^{~a_2}h_{b_1}^{~b_2}\cdots \nabla_s Q_{{a_2}{b_2}\cdots}\rt.
\ee
(1) \textbf{Scalar harmonics}: Harmonics constructed from solutions of the scalar Helmholtz equation, 
\begin{equation}
 \lp Q^{(0)} + \frac{k^2}{a^2}Q^{(0)}=0.
\end{equation}
Vector and tensor eigenfunctions constructed from the scalars are
\begin{equation}
 Q^{(0)}_a= -\frac{a}{k}h_a^{~b}\nabla_bQ^{(0)},
\end{equation}
\begin{eqnarray}
 Q^{(0)}_{ab} &=& -\left(\frac{a}{k}\right) \nabla_{\langle a}Q^{(0)}_{b \rangle}\nonumber \\
              &=& \left(\frac{a}{k}\right)^2 \nabla_{\langle a}\nabla_{b \rangle}Q^{(0)}\nonumber \\
              &=& \left(\frac{a}{k}\right)^2 h_{(a}^{~c}h_{b)}^{~d}\nabla_c\nabla_dQ^{(0)}+\fb h_{ab}Q^{(0)}.
\end{eqnarray}

(2) \textbf{Vector Harmonics}: Harmonics constructed from solutions of the vector Helmholtz equation, 
\begin{equation}
 \lp Q^{(1)}_a + \frac{k^2}{a^2}Q^{(1)}_a=0, \nabla^a Q^{(1)}_a=0.
\end{equation}
Tensor eigenfunctions constructed from the vectors are

\begin{eqnarray}
 Q^{(1)}_{ab} = -\left(\frac{a}{k}\right) \nabla_{\langle a}Q^{(1)}_{b \rangle}.
\end{eqnarray}

(3) \textbf{Tensor Harmonics}: Harmonics constructed from solutions of the tensor Helmholtz equation, 
\begin{equation}
 \lp Q^{(2)}_{ab} + \frac{k^2}{a^2}Q^{(2)}_{ab}=0, \nabla^b Q^{(2)}_{ab}=0, ~~ Q^{(2)a}_{~a}=0.
\end{equation}

\section{Relation to ordinary gauge invariant variables}
\label{ordinary}

In the coordinate based perturbation theory, we consider small fluctuations of spacetime metric about the background, which in our case is a flat FLRW metric,
\be
\bar{ds}^2 = a^2(\eta)\lt -d\eta^2 + dx^idx^i \rt,
\ee
and similar fluctuation of energy-momentum tensor about an homogeneous and isotropic perfect fluid energy-momentum tensor,
\be
\bT_{\mu\nu}=(\bmu+\bp)\bu_{\mu}\bu_{\nu}+\bp\bg_{\mu\nu},
\ee
where $\bmu(\eta)$ and $\bp(\eta)$ are energy density and pressure as observed by a comoving observer with velocity $\bu^{\mu}$:
\be
\bu_{\mu}\bu^{\mu}=-1, \quad \bu_{\mu} = \lt\frac{1}{a}, \vec{0} \rt, \quad \bu^{\mu} = \lt-a, \vec{0} \rt.
\ee

Perturbations are defined as

\ba
\dg_{\mu\nu}= a^2\lt \begin{array}{cc}
-2\phi & \pa_i\mb-\mb_i \\
\pa_i\mb-\mb_i & -2\psi\de_{ij}+2\pa_i\pa_j\me+\pa_i\me_j+\pa_j\me_i+\me_{ij} \end{array} \rt, \nonumber \\
\ea

\ba
& p(\eta,\vec{x})=\bp(\eta)+\dep(\eta,\vec{x}), \quad \mu(\eta,\vec{x})=\bmu(\eta)+\dm(\eta,\vec{x}), \nonumber \\
& \bu^{\mu}=u^{\mu}+\du^{\mu}.
\ea

From $u_{\mu}u^{\mu}=-1$,
\ba
\du^0=-\frac{\phi}{a}, \quad \du_0=-a\phi
\ea
and, 
\ba
\du_i = \pa_i\muu+\muu_i, \nonumber \\ 
\du^i = \frac{1}{a^2}\Lt \pa_i\lt \muu-a\mb\rt + \lt\muu_i+a\mb_i\rt \Rt,
\ea

where $\phi, \mb, \psi, \me, \dm, \dep$ and $\muu$ are scalar, $\mb_i,\me_i$ and $\muu_i$ are divergenceless vectors and $\me_{ij}$ is a divergenceless, traceless, symmetric tensor on the 3-hypersurface in the background spacetime. We consider only perfect fluid perturbations. Hence the anisotropic stresses are zero. All of the above variables are not invariant under infinitesimal coordinate (gauge) transformation. However, we can construct some gauge invariant variables as follows:

\ba
& \Phi = \phi +  \frac{1}{a}\lt a(\mb-\me')\rt', \quad
\Psi = \psi -  \frac{a'}{a}(\mb-\me'), \nonumber \\
& \ug = \muu - a(\mb-\me'), \quad
\mg = \dm + \bmu'(\mb-\me'), \nonumber \\
\label{ord_gi}  & \pg = \dep + \bp'(\mb-\me'), 
\ea

\ba
\mb^{\mbox{GI}}_i=\mb_i+\me_i.
\ea

$\muu_i$ and $\me_{ij}$ are gauge invariant. In this section we use the notations, $()'=\frac{d}{d\eta}$, $(\dot{)}=\frac{d}{d\bar{t}}=\bu^{\mu}\bar{\nabla}_{\mu}$, $\mh=\frac{a'}{a}$.

The expansion $\ta$ can be written as
\ba
\ta = \nabla_{\mu}u^{\mu}=\bar{\ta}+\de\ta, \quad \bar{\ta} = \frac{3\mh}{a}, \\ 
\label{deta} \de\ta = - \frac{3}{a}\lt\psi'+\mh\phi\rt + \frac{1}{a^2}\vec{\nabla}^2\lt \muu +a(\me'-\mb)\rt.
\ea

The shear $\si_{ij}$ is
\ba
 \si_{ij} &=& \pa_i\pa_j\ug-\fb\vec{\nabla}^2\ug\de_{ij} \nonumber \\
 & & + \fa\Lt\pa_i\lt\muu_j+a\mb^{\mbox{GI}}_j\rt+\pa_j\lt\muu_i+a\mb^{\mbox{GI}}_i\rt\Rt \nonumber \\
 & & +\fa\me_{ij}.
\ea
 
The vorticity $\om_{ij}$ and the vector $r_i$ are

\be
\om_{ij} = \pa_j\muu_i-\pa_i\muu_j, \quad r_i=\frac{1}{a^2}\vec{\nabla}^2\muu_i
\ee

To evaluate the spatial derivative of a scalar we note the spatial projection tensor $h_{\mu}^{~\nu}$ is

\bn
h_{\mu}^{~\nu}=\lt \begin{array}{cc}
0 & -\frac{1}{a}\pa_i(\muu-a\mb)-\frac{1}{a}(\muu_i+a\mb_i)\\
\frac{1}{a}(\pa_i\muu+\muu_i) & \de_{ij} \end{array} \rt.
\en
 
Then, 
\ba
X_i  &=& \kappa\lt \pa_i\dm + \frac{1}{a}\bmu'\du_i \rt, \nonumber \\
\label{Xi1} &=& \kappa\pa_i\lt\mg+\frac{\bmu'}{a}\ug\rt+\frac{\kappa\bmu'}{a}\muu_i.
\ea

 Similarly,
\ba
Z_i &=& \pa_i(\de\ta+\dot{\ta}\muu)+\dot{\ta}\muu_i \nonumber \\
\label{Zi1}&=& \pa_i\lt- \frac{3}{a}\lt\Psi'+\mh\Phi\rt + \frac{1}{a^2}\vn\ug -\frac{3}{2}\kappa(\bmu+\bp)\ug\rt \nonumber \\
& & \ql -\frac{3}{2}\kappa(\bmu+\bp)\ug_i. \nonumber \\
\ea

We have also used the background Friedmann equations,
\ba
\dot{\bar{\ta}}=\frac{3}{a^2}(\mh'-\mh^2)=-\frac{3}{2}\kappa(\bmu+\bp), \nonumber \\
\fb\bar{\ta}^2=\frac{3\mh^2}{a^2}=\kappa\bmu.
\ea

The expressions (\ref{Xi1}) and (\ref{Zi1}) can be further simplified using the perturbation equations used in coordinate based perturbation theory. For perfect fluid perturbations (anisotropic stresses are absent) $\Phi=\Psi$,
\ba
\label{pertm} \vn\Phi-3\mh(\Phi'+\mh\Phi)=\fa\kappa a^2\mg, \\
\label{pertu} (a\Phi)'=-\fa\kappa a^2(\bmu+\bp)\ug, \\
\label{pertp} \Phi''+3\mh\Phi'+(2\mh'+\mh^2)\Phi = \fa\kappa a^2\pg.
\ea

Using (\ref{pertu}), we obtain
\ba
\ug = \frac{a}{\mh}\lt\Phi-\zeta\rt,
\ea

where $\zeta$ is the comoving curvature perturbation,
\ba
\zeta=\fc\frac{\Phi+\mh^{-1}\Phi'}{1+w}+\Phi.
\ea

Then,\ba
\label{Xi2} X_i &=& \frac{2}{a^2}\pa_i\vn\Phi-\frac{3\mh}{a}\kappa(\bmu+\bp)\muu_i, \\
\label{Zi2} Z_i &=& \frac{1}{a\mh}\pa_i\vn\lt\Phi-\zeta\rt - \frac{3}{2}\kappa(\bmu+\bp)\muu_i.
\ea

The scalar covariant perturbations are related not only to the scalar perturbations but also to the vector perturbations of coordinate based perturbation theory because in the coordinate based perturbation theory, the 3+1 decomposition is done with respect to the world lines of the background comoving observers whereas, in covariant based theory, we use the world lines of the comoving observers of the physical spacetime.

We can define another variable $V_a=X_a-\fc\ta Z_a$, such that
\be
V_i=\frac{2}{a^2}\pa_i\vn\zeta
\ee

However this $\zeta$ is related to the $\zeta_a$ defined in (\ref{zetaa}). 

\be
\label{Wi2} W_i = \pa_i\lt\frac{\de a}{a}+\fb\bar{\ta}\muu\rt + \fb\bar{\ta}\muu_i,
\ee

\be
\frac{\de a}{a} = -\psi + \fb\int\frac{1}{a}\vn\ug d\eta.
\ee

Then using (\ref{Xi2}) and (\ref{Wi2}),
\be
\zeta_i = \pa_i \lt -\zeta + \frac{2\vn\Phi}{\kappa(\bmu+\bp)a^2} - \fb\int d\eta\mh^{-1}\vn\lt\zeta-\Phi\rt \rt.
\ee 
When spatial derivatives are small, $\zeta_i\approx-\pa_i\zeta$ and $V_i\approx-\frac{2}{a^2}\vn\zeta_i$.

The spatial curvature perturbation $\zw$ is defined as
\be
\zw = \zeta + \fc \frac{\vn\Phi}{\kappa(\bmu+\bp)a^2}
\ee

It can be shown readily that $\zw$ is related to the covariant variable,
\be
\vw_a = V_a + \fc\frac{\lp X_a}{\kappa(\bmu+\bp)} + \fc\ta r_a,
\ee
as
\ba
\label{vwzw}\vw_i = \frac{2}{a^2}\pa_i\vn\zw.
\ea 



\end{document}